\documentclass[reprint, prb,amsmath,amssymb,showpacs,superscriptaddress]{revtex4-1}
\usepackage{hyperref}
\usepackage{graphics}
\usepackage{dcolumn}
\usepackage{bm}
\usepackage{color}
\usepackage{epstopdf}
\usepackage{epsfig}
\usepackage{floatrow}
\usepackage{amsmath}
\usepackage{diagbox}

\newcommand{\half}{{\ensuremath{\frac{1}{2}}}}

\usepackage[dvipsnames]{xcolor}

\usepackage{xcolor,colortbl}

\definecolor{Gray}{gray}{0.85}
\definecolor{lightgreen}{RGB}{213,252,210}
\definecolor{lightblue}{RGB}{203,247,247}
\definecolor{lightyellow}{RGB}{255,248,230}
\definecolor{lightpurple}{RGB}{214,223,255}

\newcolumntype{a}[1]{>{\centering\let\newline\\\arraybackslash\hspace{0pt}\columncolor{lightgreen}}m{#1}}
\newcolumntype{d}[1]{>{\centering\let\newline\\\arraybackslash\hspace{0pt}\columncolor{lightblue}}m{#1}}
\newcolumntype{e}[1]{>{\centering\let\newline\\\arraybackslash\hspace{0pt}\columncolor{lightyellow}}m{#1}}
\newcolumntype{f}[1]{>{\centering\let\newline\\\arraybackslash\hspace{0pt}\columncolor{lightpurple}}m{#1}}

\newfloatcommand{capbtabbox}{table}[][\FBwidth]

\setcitestyle{numbers,square}

\begin{document}

\noindent\hrulefill
\smallskip\noindent

\title{Magnetic ground state and perturbations of the distorted kagome Ising metal TmAgGe}

\author{C. B. Larsen}
\affiliation{Laboratory for Neutron Scattering and Imaging, Paul Scherrer Institute, CH-5232 Villigen, Switzerland}

\author{D. G. Mazzone}
\affiliation{Laboratory for Neutron Scattering and Imaging, Paul Scherrer Institute, CH-5232 Villigen, Switzerland}

\author{N. Gauthier}
\affiliation{Institut Quantique, Département de physique, Université de Sherbrooke, Sherbrooke, Québec J1K 2R1, Canada}

\author{H. D. Rosales}
\affiliation{Instituto de Física de Líquidos y Sistemas Biológicos (IFLYSIB), UNLP-CONICET, Facultad de Ciencias Exactas, 1900 La Plata, Argentina}
\affiliation{Departamento de Cs. Básicas, Facultad de Ingeniería, Universidad Nacional de La Plata, C.C. 67, 1900 La Plata, Argentina}

\author{F. A. Gómez Albarracín}
\affiliation{Instituto de Física de Líquidos y Sistemas Biológicos (IFLYSIB), UNLP-CONICET, Facultad de Ciencias Exactas, 1900 La Plata, Argentina}
\affiliation{Departamento de Cs. Básicas, Facultad de Ingeniería, Universidad Nacional de La Plata, C.C. 67, 1900 La Plata, Argentina}

\author{J. Lass}
\affiliation{Laboratory for Neutron Scattering and Imaging, Paul Scherrer Institute, CH-5232 Villigen, Switzerland}
\affiliation{Department of Physics, Technical University of Denmark, DK-2800 Kongens Lyngby, Denmark}

\author{X. Boraley}
\affiliation{Laboratory for Neutron Scattering and Imaging, Paul Scherrer Institute, CH-5232 Villigen, Switzerland}

\author{S. L. Bud'ko}
\affiliation{Ames National Laboratory and Department of Physics and Astronomy, Iowa State University, Ames, Iowa 50011, USA}

\author{P. C. Canfield}
\affiliation{Ames National Laboratory and Department of Physics and Astronomy, Iowa State University, Ames, Iowa 50011, USA}

\author{O. Zaharko}\email{oksana.zaharko@psi.ch}
\affiliation{Laboratory for Neutron Scattering and Imaging, Paul Scherrer Institute, CH-5232 Villigen, Switzerland}

\date{\today}

\begin{abstract}
 We present the magnetic orders and excitations of the distorted kagome intermetallic magnet TmAgGe. Using neutron single crystal diffraction we identify the propagation vectors $\bf{k}$ = ({\half}~0~0) and $\bf{k}$ = (0~0~0) and determine the magnetic structures of the zero-field and magnetic field-induced phases for $H$ along the $a$ and $[-1 1 0]$ crystal directions. We determine the experimental magnetic field- temperature ($H$, $T$)-phase diagram and reproduce it by Monte-Carlo simulations of an effective spin exchange Hamiltonian for one distorted kagome layer. Our model includes a strong axial single-ion anisotropy and significantly smaller exchange couplings which span up to the third-nearest neighbours within the layer. Single crystal inelastic neutron scattering (INS) measurements reveal an almost flat, only weakly dispersive mode around 7 meV that we use alongside bulk magnetization data to deduce the crystal-electric field (CEF) scheme for the Tm$^{3+}$ ions. Random phase approximation (RPA) calculations based on the determined CEF wave functions of the two lowest quasi-doublets enable an estimation of the interlayer coupling that is compatible with the experimental INS spectra. No evidence for low-energy spin waves associated to the magnetic order was found, which is consistent with the strongly Ising nature of the ground state.
\end{abstract}

\maketitle

\section{Introduction}{\label{Sec1}}

\begin{figure}
    \centering
    \includegraphics[width=8.6cm]{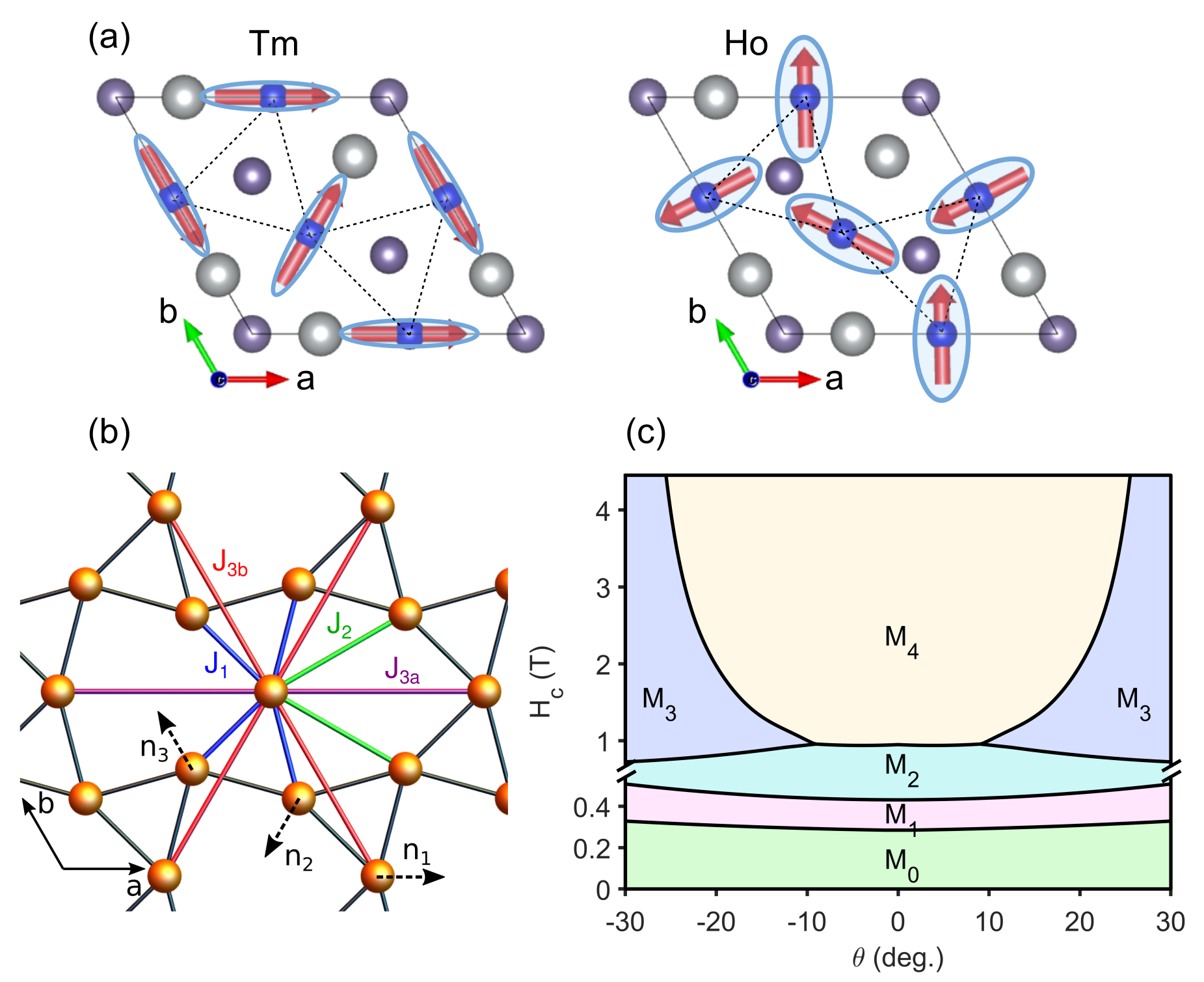}
    \caption{(a) Comparison of the local CEF anisotropy directions of TmAgGe and HoAgGe depicted as blue ellipses. (b) A single layer of the distorted kagome structure with $J_1$, $J_2$, $J_{3a}$ and $J_{3b}$ being exchange interactions used in the model represented by Eq.~(\ref{eq:Hamil}). $n_i$ ($i = 1,2,3$) indicate the three easy crystal electric field (CEF) axes.(c) Schematic phase diagram of TmAgGe based on  magnetization data ~\cite{morosan2005}. The angle $\theta$ denotes the departure of the magnetic field from the crystallographic $a$-axis within the hexagonal $ab$-plane, H$_c$ is a critical field value. }
    \label{fig:phd}
\end{figure}

The interaction between itinerant conduction electrons and localised magnetic moments gains presently renewed interest. This is driven by the discovery of unconventional magnetoresistive properties~\cite{surgers2017, kurumaji2019} emerging when itinerant electrons travel through the lattice of non-collinear magnetic moment arrangements. The shape and topology of the Fermi surface is crucial in mediating the magnetic interactions between localised moments, but also the conduction electrons adjust when complex magnetic order sets in. They pick up a Berry phase from adjacent noncollinear spins, which induces a topological Hall effect as the example of Mn$_3$Ge shows~\cite{nayak2016}. The interplay of the two electron subsystems - itinerant and localised - could lead to nesting of the Fermi surface, formation of multi-k magnetic orders and other exotic behaviours. The case with localised spins on a geometrically frustrated lattice is especially interesting~\cite{batista2008, sengupta2017}. For example, in the Kagome system Gd$_3$Ru$_4$Al$_{12}$ chiral spin fluctuations influence transport anomalies even in the paramagnetic regime~\cite{kolincio2023}.\\
    Therefore, we focus on the family of magnetically frustrated rare-earth RAgGe compounds which possess hexagonal distorted kagome layers spanned by the R$^{3+}$ ions. The series shows complex transport properties~\cite{morosan2004} signifying a cascade of magnetic orders and metamagnetic transitions as a function of temperature and magnetic field.\\
    The magnetic properties of RAgGe are governed by the local single-ion anisotropy and magnetic exchanges that are geometrically frustrated. Their interplay varies for different magnetic ions which we exemplify here comparing HoAgGe and TmAgGe. The R$^{3+}$ ions obey an orthorhombic $C_{2v}$ symmetry which is axial and confined within the kagome layer. Whereas the Ho anisotropy axis is orthogonal to the 2-fold axis, it is along the axis for Tm (Fig.~\ref{fig:phd}a).  The overall hexagonal $P \bar{6} 2 m$ crystal symmetry leads to a rotation of the anisotropy axis of the three neighboring R$^{3+}$ ions by 60$^\circ$ relative to each another. The magnetic exchange $J_1$ is predominantly ferromagnetic at the first neighbour distance for both compounds. The further-neighbour in-plane couplings $J_2$, $J_3$ are defined in Fig.~\ref{fig:phd}b and Table~\ref{table:Js}. While also sizeable, they are predominantly antiferromagnetic~\cite{zhao2020, goddard2007}.\\ 
    In HoAgGe~\cite{zhao2020} the interplay between the single-ion anisotropy and nearest-neighbor ferromagnetic exchange establishes a magnetic structure where the magnetic moments  point 'in' or 'out' of  edge-sharing $J_1$-triangles in a Kagome layer (Fig.~\ref{fig:phd}b). These magnetic states obey the 'two-in-one-out' or 'one-in-two-out' Kagome spin-ice rule. The Ising anisotropy allows the application of an effective pseudospin-1/2 classical Ising model to this metallic system.\\ 
    In TmAgGe the Kagome spin-ice rule is disregarded on the $J_1$-triangles, as the CEF easy axes of the R$^{3+}$ ions is perpendicular to the easy axis in HoAgGe (Fig.~\ref{fig:phd}c). The microscopic interactions lead to a cascade of metamagnetic transitions when a magnetic field is applied within the kagome layer (Fig.~\ref{fig:phd} c)~\cite{morosan2005, goddard2007}. These transitions are thought to arise via spin flip transitions along the local easy-axes. This is supported by a phenomenological triple coplanar Ising-like model~\cite{morosan2005} that accounts for the spin flips and successfully reproduces the macroscopically determined ($H$, $T$)-phase diagram. A microscopic Hamiltonian proposed by Goddard $et$ $al.$ \cite{goddard2007} predicts strong single-ion anisotropy ($A$= -4.6(1) K) and almost equal size first nearest-neighbor couplings ($J_1$= -0.064(3) K and $J_2$ = 0.054(3) K). Yet the experimental data could be reproduced only when some $J_1$ couplings are suppressed.\\
    In this paper we present a neutron single crystal diffraction (Sec.~\ref{Sec2_ND}) and inelastic neutron scattering (Sec.~\ref{Sec2_INS}) study on TmAgGe and model these results (Secs.~\ref{Sec3}, \ref{Sec4}) to determine the magnetic structures and the CEF scheme of the Tm$^{3+}$ ions. Monte-Carlo simulations (Sec.~\ref{Sec5}) of a Hamiltonian including strong single-ion anisotropy and Heisenberg-type bilinear exchanges quantify couplings within a distorted kagome layer. RPA calculations  (Sec.~\ref{Sec6}) based on the wavefunctions of the two lowest CEF levels validate our understanding of the magnetic properties of TmAgGe. 

\section{Experimental Details}{\label{Sec2}}

\subsection{Sample synthesis}

 Single crystals of TmAgGe were grown in a manner similar to that outlined in Refs.~\citenum{morosan2004} and~\citenum{morosan2005}.  Specifically, elemental Tm, Ag and Ge were placed in a fritted crucible set~\cite{canfield2016}, sold by LSP ceramics~\cite{lsp} as a Canfield Crucible Set (CCS), in the atomic ratios of Tm$_9$Ag$_{68}$Ge$_{23}$.  The CCS was sealed into an amorphous silica ampule under a partial pressure of Ar ($\sim$ 1/6 atm)~\cite{canfield2020}.  The ampule was then heated to 1190$^\circ$C over 10  hours, held at 1190$^\circ$C for 10 hours and then cooled to 850$^\circ$C over 400 hours. After reaching 850$^\circ$C the ampule was removed from the furnace and placed into the rotor of a centrifuge that was used to provide an enhanced, local acceleration that forced excess liquid through the frit, leaving TmAgGe single crystals on the growth side crucible.  Single crystals, with mirrored facets and masses in excess of 0.2\,g were grown.
 
\subsection{Neutron diffraction}{\label{Sec2_ND}}

\begin{figure*}
    \centering
    \includegraphics[width = 14cm]{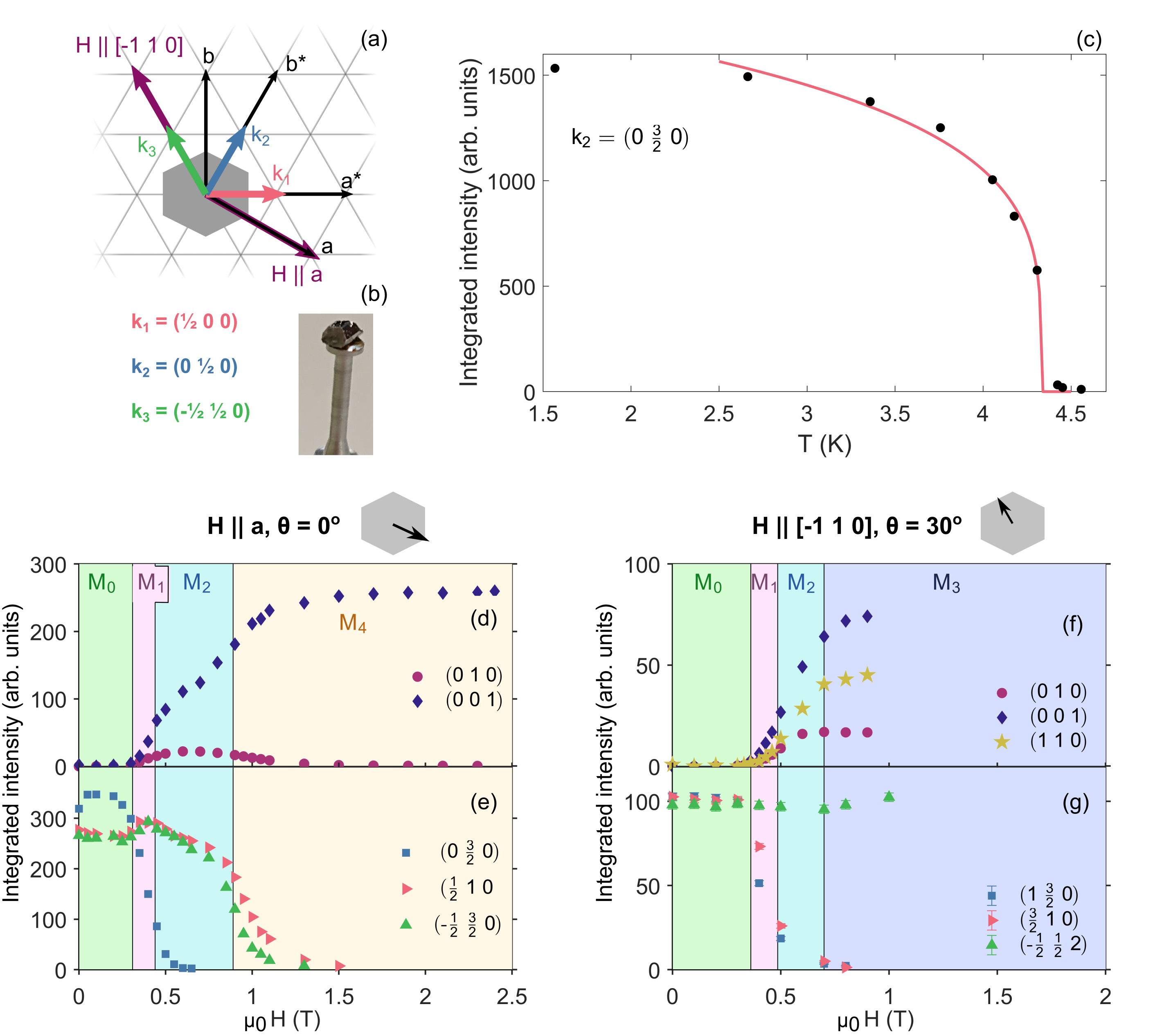}
    \caption{(a) Schematic view of the $\bf{k}_1$=({\half}~0~0), $\bf{k}_2$=(0~{\half}~0), and $\bf{k}_3$=({-\half}~{\half}~0) propagation vectors in relation to the in-plane direct-space ($a$, $b$) and reciprocal-space ($a^*$, $b^*$) lattice vectors. (b) Example of a $\sim$0.2\,g TmAgGe single crystal used for neutron diffraction. (c) Zero-field temperature dependence of the intensity of the (0~${\ensuremath{\frac{3}{2}}}$~0)  magnetic reflection, alongside a fitted line (red). (d-g) Field-dependence of the integrated intensities of select magnetic peaks (left scale) while ramping the magnetic field up with the two principal field configurations. The data were collected at temperatures of 2\,K ($\theta$=0$^\circ$) and 1.7\,K ($\theta$=30$^\circ$).
    }
    \label{fig:fielddep}
\end{figure*}

     Single-crystal neutron-diffraction (SND) experiments at zero field and under field were performed with the single-crystal neutron diffractometer Zebra at the Swiss Neutron Spallation Source SINQ, Paul Scherrer Institut (PSI), Switzerland. The experiments were carried out with small $\sim$0.2\,g TmAgGe single crystals to limit absorption and extinction effects. A neutron wavelength of 1.383\,\AA\ was used for all measurements, offering a compromise between flux and accessibility in reciprocal space. In zero field, Zebra was operated in 4-circle mode with a Joule-Thomson CCR3 cooling machine. In-field data were collected within a lifting-arm normal-beam geometry where the crystals were inserted in vertical magnets. The magnetic states were investigated under two principal field configurations: $H \| a$ corresponding to $\theta = 0^\circ$ and $H \|$ [-1 1 0] (equivalent to $a^*$ and [1 2 0] used in Ref.~\citenum{morosan2005}) where $\theta = 30^\circ$, see Fig.~\ref{fig:fielddep}a.

    In accordance to powder neutron diffraction results~\cite{baran2009}, the zero field antiferromagnetic propagation vector is $\bf{k}$=({\half}~0~0) at temperatures below $T_N$ = 4.3\,K. Using SND we observed the three arms  $\bf{k}_1$=({\half}~0~0), $\bf{k}_2$=(0~{\half}~0), and $\bf{k}_3$=({-\half}~{\half}~0). Figure~\ref{fig:fielddep}c depicts the temperature dependent integrated intensity of the (0~$\frac{3}{2}$~ 0) magnetic reflection. The red line signifies a fit $I(T) \propto  [(T_N - T) / T_N]^{2 \beta}$. The  N\'eel temperature is refined to $T_N = 4.33(1)$\,K and the critical exponent $\beta$ is 0.12(1). The fitted critical exponent is in agreement with the theoretical value of 1/8 which is expected for 2D Ising systems~\cite{kadanoff1966}. 
    
    The application of a magnetic field affects the magnetic intensities in the three $\bf{k}$ arms, and leads to the emergence of an additional $\bf{k}_0$=(0~0~0) order. Figures~\ref{fig:fielddep}d-g summarise the field dependence of the integrated intensities for selected reflections representing all four propagation vectors alongside region boundaries from critical field measurements~\cite{morosan2005}. For $H \| a$ (Fig.~\ref{fig:fielddep} (d, e)), a clear correspondence between the integrated neutron intensities and critical regions is found. In the M$_0$ phase the net macroscopic magnetic moment is known to be zero~\cite{morosan2005}, which is reflected in the SND data where only reflections of the three $\bf{k}_{1-3}$ propagation vectors exhibit finite intensities. The onset of the M$_1$ region at $\sim$0.31\,T is marked by an increasing net magnetic moment along the $a$-direction in the bulk measurements. This coincides with the appearance of the $\bf{k}_0$ order and the initial decay of the intensity of the $\bf{k}_2$ reflection. The M$_1$ region is thought to represent, at the measurement temperatures of $\sim$2\,K, a transitory metastable region and no plateau is reported in the magnetization measurements for this field range~\cite{morosan2005}. The M$_2$ region begins at $\sim$0.44\,T, where the $\bf{k}_2$ intensity fully disappears and the $\bf{k}_1$ and $\bf{k}_3$ intensities start to decline. The magnetic $\bf{k}_0$-intensities exhibit a different behaviour for different reflections; The intensity of the (0~0~1) reflection steadily increases, while the (0~1~0) reflection peaks in the M$_2$ region and eventually disappears in the M$_4$ region. For $H\|$[-1~1~0], the     $\bf{k}_0$ intensities monotonically increase as a function of increasing field strength until they saturate in the M$_3$ phase (Fig.~\ref{fig:fielddep} (f, g)). The magnetic intensities of the $\bf{k}_1$ and $\bf{k}_2$ reflections decrease at the onset of the M$_1$ region and vanish in the M$_3$ phase. The intensity corresponding to the third propagation vector $\bf{k}_3$, which is parallel to the field direction, maintains finite intensity at the highest measured field of 6\,T. We note that for the both field configurations the intensity remains in the $\bf{k}_{1-3}$ propagation vectors that have the largest component parallel to the field direction.

\subsection{Inelastic neutron scattering}{\label{Sec2_INS}}

\begin{figure}
    \centering
    \includegraphics[width = 8.6cm]{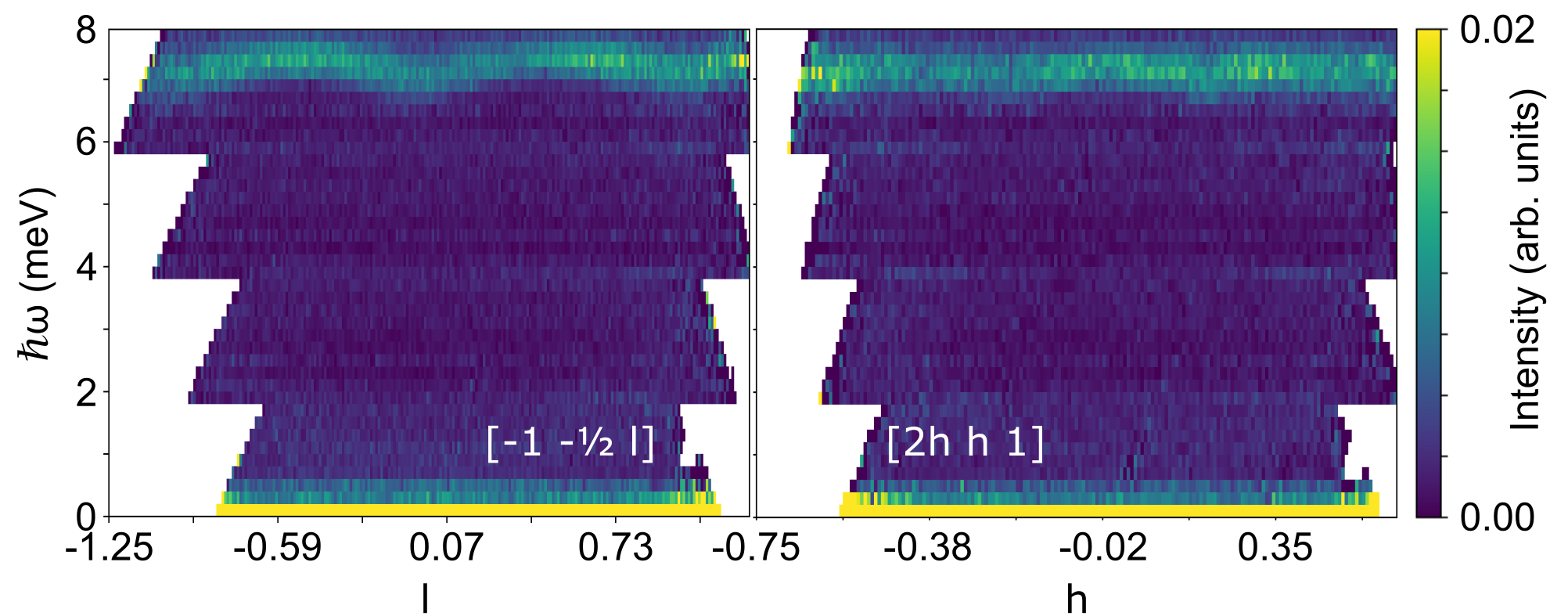}
    \caption{Zero-field 2\,K QE-cuts along the [-1 -$\frac{1}{2}$ $l$] and [$2h$ $h$ 1]-directions at the CAMEA multiplexing spectrometer. The data are shown in Q-steps of 0.01 rlu and energy steps of 0.2 meV. The integration width corresponds to 0.2 rlu perpendicular to Q.}
    \label{fig:cameamay}
\end{figure}

Inelastic neutron scattering (INS) data were collected on the CAMEA multiplexing neutron spectrometer at PSI, Switzerland\cite{marko2018, lass2020}. In a first zero-field experiment, a $\sim$2\,g single-crystal TmAgGe sample was inserted into an orange cryostat. At each incident energy ranging from 5-11\,meV the sample was rotated over 100$^\circ$ in one degree steps acquiring data for about 60 s in two detector settings to gain a broad overview of the inelastic spectrum (see Fig. \ref{fig:cameamay}). The main characteristic feature of the spectrum is a weakly dispersing mode at $\sim$7\,meV, which was observed to persist above the ordering temperature. No low-energy spin-wave excitations were observed.

A subsequent in-field experiment was pursued to investigate the 7\,meV CEF mode. To improve the signal, we co-aligned two TmAgGe single crystals with a combined sample mass of $\sim$3\,g and a mosaicity of 1$^\circ$. The crystals were aligned with the $a$-direction vertical, corresponding to the $\theta = 0^\circ$ field configuration, and installed into a vertical 11\,T cryomagnet. Quasi-elastic spectra were first collected at a base temperature 2\,K and at fields of ~0\,T and 0.65\,T using two incident energies close to E$_i$~=~5 meV. The resulting elastic energy and Q-resolution was of the order of $\Delta E$~$\approx$~0.2\,meV and 0.05 \AA$^{-1}$, respectively.
The CEF mode was investigated using a combination of three incident energies around E$_i$ = 11.3 ($\Delta E$ $\approx$ 0.5\,meV at and excitation energy of 7\,meV) under several different field and temperature conditions ranging from 2-25\,K and 0-8\,T. The slightly different energy settings provided coverage of all instrumental black spots and optimized the signal to noise ratio along the energy. 

\section{Determination of magnetic structures}{\label{Sec3}}

\begin{figure}
    \centering
    \includegraphics[width = 8.6cm]{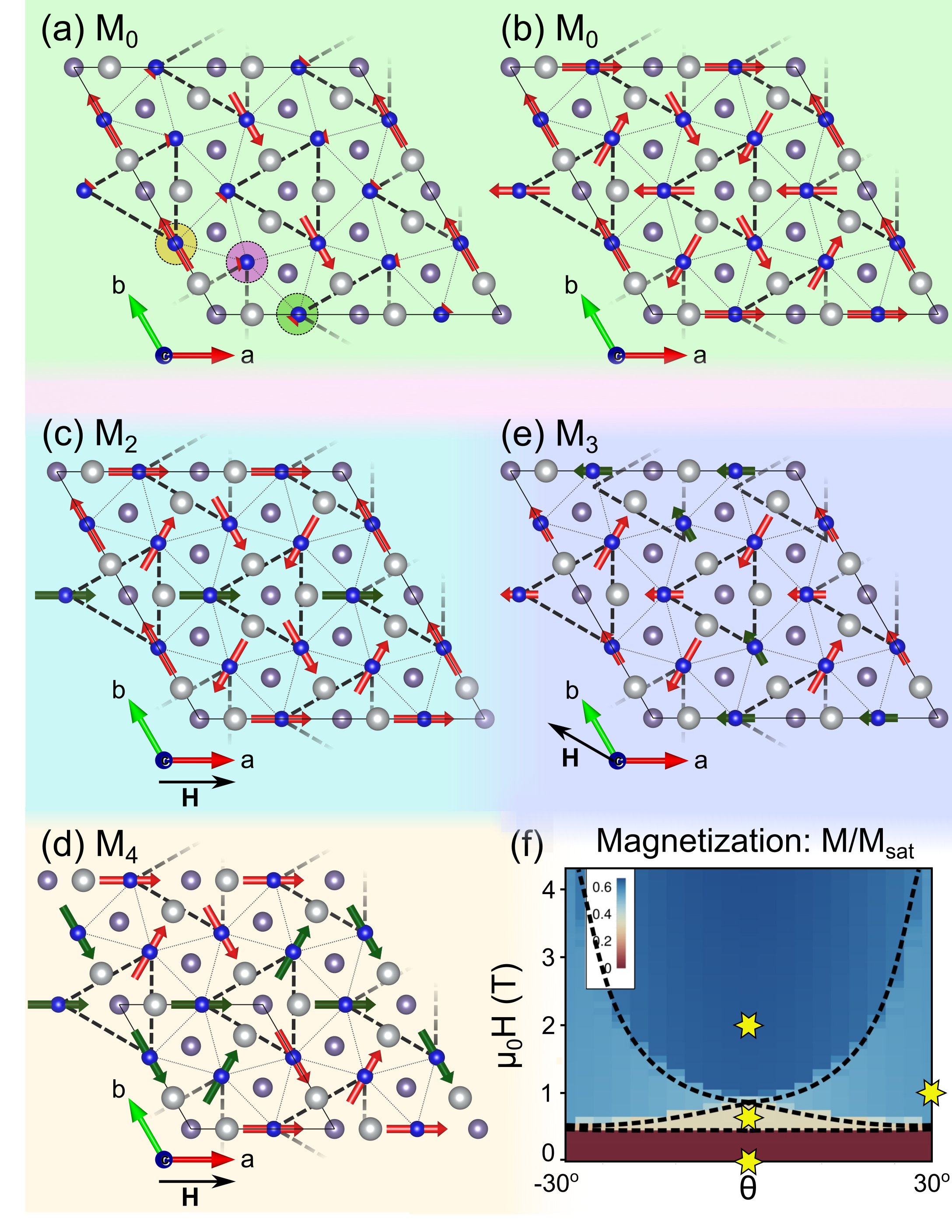}
    \caption{In-plane view of the magnetic structures determined by single crystal neutron diffraction at 2\,K and at various field configurations. The nearest-neighbour moments are connected by dotted grey lines, while the next-nearest-neighbours are connected by dashed black lines. Moments that are flipped in comparison to the zero-field multi-$k$ structure have been drawn in a dark green colour. (a) Best single-$k$ ($\bf{k}_1$) and (b) multi-$k$ structures ($\bf{k}_1$+$\bf{k}_2$+$\bf{k}_3$) obtained from zero-field data (M$_0$). The three colored circles in the lower left corner indicate the positions of the three magnetically different Tm sites; Tm1 (green), Tm2 (yellow), and Tm3 (red). (c) Multi-$k$ solution ($\bf{k}_0$+$\bf{k}_1$+$\bf{k}_3$) obtained with $H \| a$ and $\mu_0 H$ = 0.65\,T, corresponding to the M$_2$ region of the phase diagram. (d) Single-$k$ solution ($\bf{k}_0$) in the M$_4$ phase with $H \| a$ and $\mu_0 H$ = 2\,T. (e) Multi-$k$ solution ($\bf{k}_0$+$\bf{k}_3$) from data measured with $H \| [-1 1 0]$ and $\mu_0 H$ = 1\,T (M$_3$). Plots of the magnetic structures have been generated using the {\it VESTA} visualization program~\cite{momma2011}. (f) H$_c$-$\theta$ phase diagram computed from Monte Carlo simulations. The colour scale of the phase diagram expresses the magnetization in units of the saturation magnetization. Yellow stars indicate field configurations where the magnetic structure has been characterized with SND.}
    \label{fig:refinedstruc}
\end{figure}

\subsection{Zero-field magnetic structure}

    Zero-field data sets were collected for all three arms of the $\bf{k}$=({\half}~0~0) propagation vector at a temperature of 2\,K. As an initial approach to resolve the magnetic structure, separate single-$k$ refinements were carried out for each data set by iterating through the possible magnetic space group (MSG) solutions which were derived from the \emph{k-SUBGROUPSMAG} program~\cite{perezmato2015} of the Bilbao Crystallographic server~\cite{aroyo2006, aroyo2006v2, aroyo2011}. These solutions are listed in Fig.~\ref{fig:MSG_singlek} b. For all three data sets, the best agreements in terms of goodness-of-fit and fit-parameter uncertainty were found with the $P_Anc2$ MSG. This model enforces a basal-plane structure with two orbits; the (x,0,0) Tm1 moments are restricted to the CEF-anisotropy direction, while the linked Tm2 (0,x,0) and Tm3 (-x,-x,0) moments can deviate from their respective anisotropy directions. The refined magnetic solutions were consistent among all three arms, each structure containing a partial magnetic ordering with only one site displaying a moment-amplitude larger than its uncertainty, see Fig.~\ref{fig:refinedstruc}a and Table~\ref{tab:refinement_summary}, where moment values in the range of 6.1(1)-6.8(1) $\mu_B$ are reported for three different solutions for the sites with finite moment-values. 
    
    Since such order is unlikely, multi-$k$ solutions have been explored. As shown in Fig.~\ref{fig:MSG_multik}b multi-$k$ structures increase the solution space from 17 to 26 possible MSGs. Using the simulated annealing option of \emph{FullProf}~\cite{fullprof}, it was possible to refine a single-parameter equal-moment magnetic structure with 6.3(1) $\mu_B$ per site that is simultaneously in agreement with the data sets of all three arms and which has goodness-of-fit parameters similar to the single-$k$ refinements, see Fig.~\ref{fig:refinedstruc}a and Table~\ref{tab:refinement_summary}. The simulated annealing procedure was initially performed in the lowest-symmetry $P1$ MSG, while gradually decreasing the number of free parameters based on the fit performance. The $Pm'$ subgroup from Fig.~\ref{fig:MSG_multik}a is the highest-symmetry space group that the refined solution maps onto. This subgroup has 24 free parameters and allows unconstrained in-plane arrangement of the 12 moments in the magnetic cell. Thus, the symmetry-derived solution is less restrictive than the final solution, where the moments are confined to the local CEF-directions, as described in Refs.~\citenum{morosan2005} and \citenum{goddard2007}.
    
    In an attempt to understand whether TmAgGe can be a candidate for topological magnetism, we calculate the vector chirality for each given solution. Interplay between conduction electrons and a chiral magnetic structure can result in the emergence of topological properties such as a spin-dependent Berry phase. Hence, the spin chirality serves as an indicator of whether the local magnetic structure or its fragments could be a source of nontrivial topology and anomalous Hall effect. We calculate the vector chirality, ${\bf{\kappa}}$, in the following way~\cite{hagihala2019}:
    \begin{equation}
         {\bf{\kappa}} = \sum_{ijk} \frac{2}{3 \sqrt{3}} \epsilon_{ijk}(\hat{{\bf{m}}}_i \times \hat{{\bf{m}}}_j) 
    \end{equation}
    where $i$, $j$, $k$ refer to the three different sites in the nearest-neighbour triangles, $\epsilon_{ijk}$ is the Levi-Civita symbol and $\bf{\hat{m}_i}$ is the magnetic moment at site $i$, normalized to 1. The vector chirality $\bf{\kappa}$ is a $3\times1$ vector, which will only have a finite component in its third row ($\kappa_c$) for a fully basal plane structure. In the case of the zero-field multi-$k$ structure, 2/3 of the nearest-neighbour triangles have a $\kappa_c$ value of -1/3, while the remaining 1/3 have a $\kappa_c$ value of 1. As such, the total chirality of the structure sums up to 0, implying that topological phenomena are either non-existent or restricted to local structures in the zero-field state.

\subsection{Field along $a$-direction}
    
    A data set containing reflections from the $\bf{k}_0$, $\bf{k}_1$, and $\bf{k}_3$ propagation vectors was collected with $H \| a$ at $\mu_0 H$~=~0.65\,T, where the intensity of the (0~1~0) reflection is maximal (Fig.~\ref{fig:fielddep}d). Both cases of several simultaneous single-$k$ structures and a single multi-$k$ structure with all three propagation vectors participating in the ordering were considered. The possible magnetic space groups derived from the single-$k$ $\bf{k}_0$ and multi-$k$ set of propagation vectors are listed in Figs.~\ref{fig:MSG_singlek}a and \ref{fig:MSG_multik}b, respectively. Similarly to the zero-field case, single-$k$ refinements of the $\bf{k}_1$ and $\bf{k}_3$ data sets resulted in partially ordered magnetic structures belonging to the $P_Anc2$ MSG. The same was found to be the case for the single-$k$ treatment of the $\bf{k}_0$ data set, where the best refinement was obtained with the $Am'm2'$ MSG which again resulted in a partially ordered magnetic structure. Both the $\bf{k}_0$, $\bf{k}_1$, and $\bf{k}_3$ solutions had finite moments on only 2/3 sites with moment amplitudes in the range of 4.6(2)-4.9(1)~$\mu_B$. A multi-$k$ refinement with all three data sets resulted in a single-parameter magnetic structure with a moment-value of 6.7(1)\,$\mu_B$ per site and similar goodness-of-fit values to the single-$k$ cases. As for the zero-field case, we give preference to the multi-$k$ structure, which has fewer free parameters and shows a better agreement with bulk measurements. For the single-$k$ $k_0$ structure we calculate a net magnetic moment of 1.7(2) $\mu_B$/Tm along the field direction. In the multi-$k$ structure the magnetic moment is 2.23(3)$\mu_B$/Tm, cf.~Table~\ref{tab:refinement_summary}, which is closer to bulk magnetization measurements, reporting a value of $\sim$ 2.36 $\mu_B$/Tm~\cite{morosan2005}.  The refined multi-$k$ structure is shown in Fig.~\ref{fig:refinedstruc} c. Compared to the zero-field multi-$k$ solution some of the Tm1 (x,0,0) moments are flipped, such that all Tm1 moments now point in the same direction (along the field-direction \emph{and} the local CEF direction). Despite this, the highest-symmetry space group that this solution maps onto is still the $Pm'$ MSG (Fig.~\ref{fig:MSG_multik} b). The overall vector chirality sums up to zero, indicating that the small field did not induce any topological changes.

    Another data set was collected with $H \| a$ and $\mu_0 H$~=~2\,T, where only $\bf{k}_0$ order remains. This field configuration corresponds to the $M_4$ phase in Fig.~\ref{fig:fielddep} and has been speculated to correspond to a crystal-field limited saturated paramagnetic state (CL-SPM) in literature\cite{morosan2005, goddard2007}. The best refinement was obtained with the basal-plane $Pm'$ MSG, though it was found that the number of free parameters could be further reduced from three that is given by symmetry down to one, corresponding to a moment-value of 6.1(1)\,$\mu_B$ per site. The obtained magnetic structure is shown in Fig.~\ref{fig:refinedstruc}d.  Unlike the other structures, this magnetic phase does exhibit a finite overall chirality, with $\kappa_c$ = $-1/3$. As such, this phase would be interesting for further transport studies to establish whether topological properties can be established. The net magnetic moment along the field direction is 3.6(1) $\mu_B$/Tm, which is reduced compared to the bulk value of $\sim$~4.6 $\mu_B$~\cite{morosan2005}.

\subsection{Field along [-1 1 0] direction}

    A data set containing reflections from the $\bf{k}_0$ and $\bf{k}_3$ propagation vectors has been collected with field along [-1~1~0]. The data set was measured at base temperature and at 1\,T, corresponding to the M$_3$ region of the phase diagram, which is accessible only by rotating the field away from the $a$-direction. The single-$k$ $\bf{k}_3$ solution spans only the Tm3 sublattice and is strongly amplitude-modulated, having moment values of 7.9(1)\,$\mu_B$ per Tm3 site, the $\bf{k}_0$ model comprises the two other Tm1, Tm2 sublattices, with moment values of 5.9\,$\mu_B$ determined by a common parameter. Single-$k$ and multi-$k$ attempts also here return similar goodness-of-fit values. 
    For the multi-k solution, The Tm3 site featuring a CEF anisotropy direction normal to the field, contributes to the $\bf{k}_3$ intensities and has the largest moment value of 8.2(1)\,$\mu_B$/Tm, while two other sites have moment values of 5.3(2)\,$\mu_B$/Tm. Thus, compared to the multi-$k$ solutions obtained in other field conditions, this solution has two free parameters and exhibits a slight amplitude-modulation. Allowing the moments to diverge from the CEF directions did not significantly improve the goodness-of-fit or increase the Tm1 and Tm2 moment amplitudes, indicating that the observed amplitude-modulation is not an artifact of the imposed magnetic model. We presume the rather high moment value at the Tm3 site is due to deficient diffraction data collected in the non-optimal geometry in the magnet. Similar to the other two multi-$k$ solutions, this model maps onto the $Pm'$ MSG. The chirality of the proposed multi-k solution is calculated as -1/3. Of all the investigated phases in the $H_c-\theta$ phase diagram, the  M$_3$ region is hence the only region with a magnetic structure solution that is both multi-k and exhibits a finite chirality summed over the entire structure. This region is therefore especially interesting for further investigations of possible topological effects.

\section{Crystal-field level scheme}{\label{Sec4}}

\begin{figure*}
    \centering
    \includegraphics[width = 0.9 \linewidth]{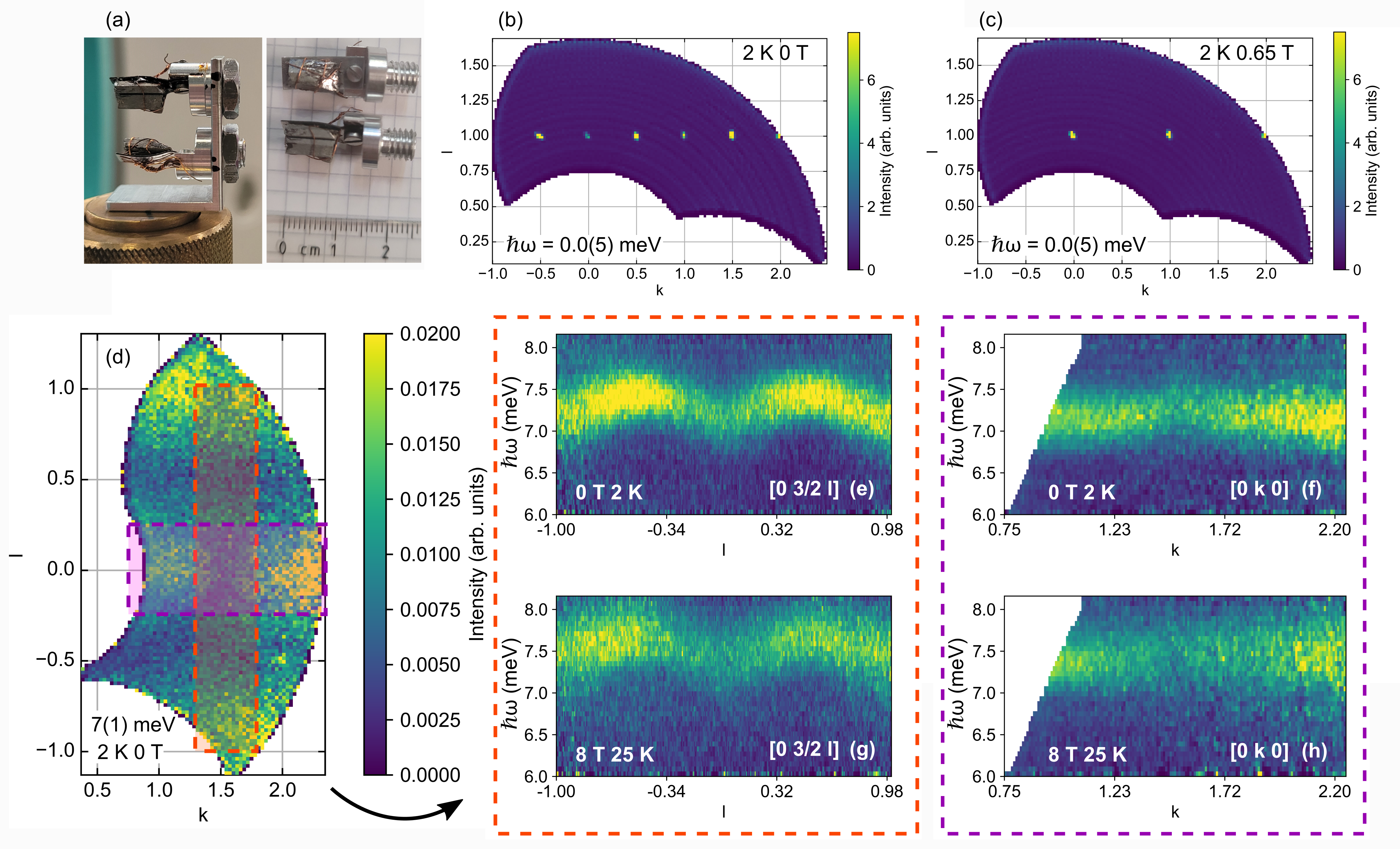}
    \caption{(a) Pictures of the two TmAgGe samples used for the INS in-field experiment. The crystals were aligned with the $a$-axis along the field direction. (b-c) Quasi-elastic scattering of the ($0kl$)-plane measured at 2\,K and for fields of 0\,T and 0.65\,T, respectively, with intensities integrated over $\hbar \omega$~= -0.05-0.05\,meV. (d) ($0kl$) scattering plane measured at zero field and a temperature of 2\,K, with intensities integrated over $\hbar \omega$~=~6-8\,meV. The vertical and horizontal boxes indicate the integrated areas for Q$\omega$-cuts (panels (e-h)) along $l$ and $k$, respectively. For Panels (b-d) we used a Q-binning of 0.03 \AA$^{-1}$. (e-h) Q$\omega$-cuts along [0~$\frac{3}{2}$~$l$] and [0~$k$~0] obtained at field and temperature combinations of 0\,T, 2\,K and 8\,T, 25\,K. The data is plotted for every 0.01 rlu along Q and 60 $\mu eV$ along the energy.}
    \label{fig:cefoverview}
\end{figure*}

    A common feature of the magnetic structures in TmAgGe is the strong alignment of the moments along the local CEF anisotropy directions, which were initially identified from bulk measurements~\cite{morosan2004, goddard2007}. In Ref. \citenum{goddard2007}, $H \| c$ magnetization data were used to determine the CEF Hamiltonian:
    \begin{equation}
         \hat{H} = \sum_{n = 2,4,6} \sum_{m=0}^n B_{n}^m \hat{O}_{n}^m - g_J \mu_0\mu_B {\bm H} \cdot \sum_i {\bm \hat{J}}_i
    \end{equation}
    where $B_{n}^m$ and $\hat{O}_{n}^m$ are Stevens coefficients and operators,respectively~\cite{stevens1952}. The Hamiltonian has been evaluated for the orthorhombic $C_{2v}$ ($2mm$) symmetry of the Tm-sites, giving raise to four finite Stevens coefficients,  $B_2^0$~=~1300\,mK, $B_4^0$~=~-3.1\,mK, $B_4^2$~=~19\,mK, and $B_6^0$ = 0.0068\,mK. In our current investigation, we aim to further improve and refine the CEF Hamiltonian by adding insight via single-crystal inelastic neutron scattering. For all CEF schemes presented in this section, we choose the local coordinate frame such that $x$ is along the 2-fold axis of the Tm$^{3+}$ ion, $z$ is along the 6-fold axis of the crystal, $y$ completes the right-handed coordinate system.

    Fig.~\ref{fig:cefoverview} summarizes the main features of the data collected during the in-field inelastic neutron scattering study of single crystaline TmAgGe (Panel (a)). Panels (b) and (c) show the quasielastic scattering results at base temperature and fields of 0 and 0.65\,T  with $H \| a$. We do not find any signatures of low-energy spin waves which is associated to the strong Ising nature of the Tm$^{3+}$ moments and explained in the Discussion (Sec.~{\ref{Sec7}}). 
    
    Fig.~\ref{fig:cefoverview}d shows the $0kl$ scattering plane at excitations energies around the $\sim$~7\,meV CEF mode. Panels (e)-(f) display energy cuts along $k$- and $l$-directions at the two extreme ends of the investigated fields and temperatures, $ie$ T~=~2\,K and $\mu_0 H$~=~0\,T alongside T~=~25\,K and $\mu_0 H$~8\,T. We find a dispersion that is strongest along the $l$-direction, and that field and/or temperature cause a slight shift of the dispersion and a decrease of intensity. 
    
\begin{figure}
        \centering
        \includegraphics[width=8.6cm]{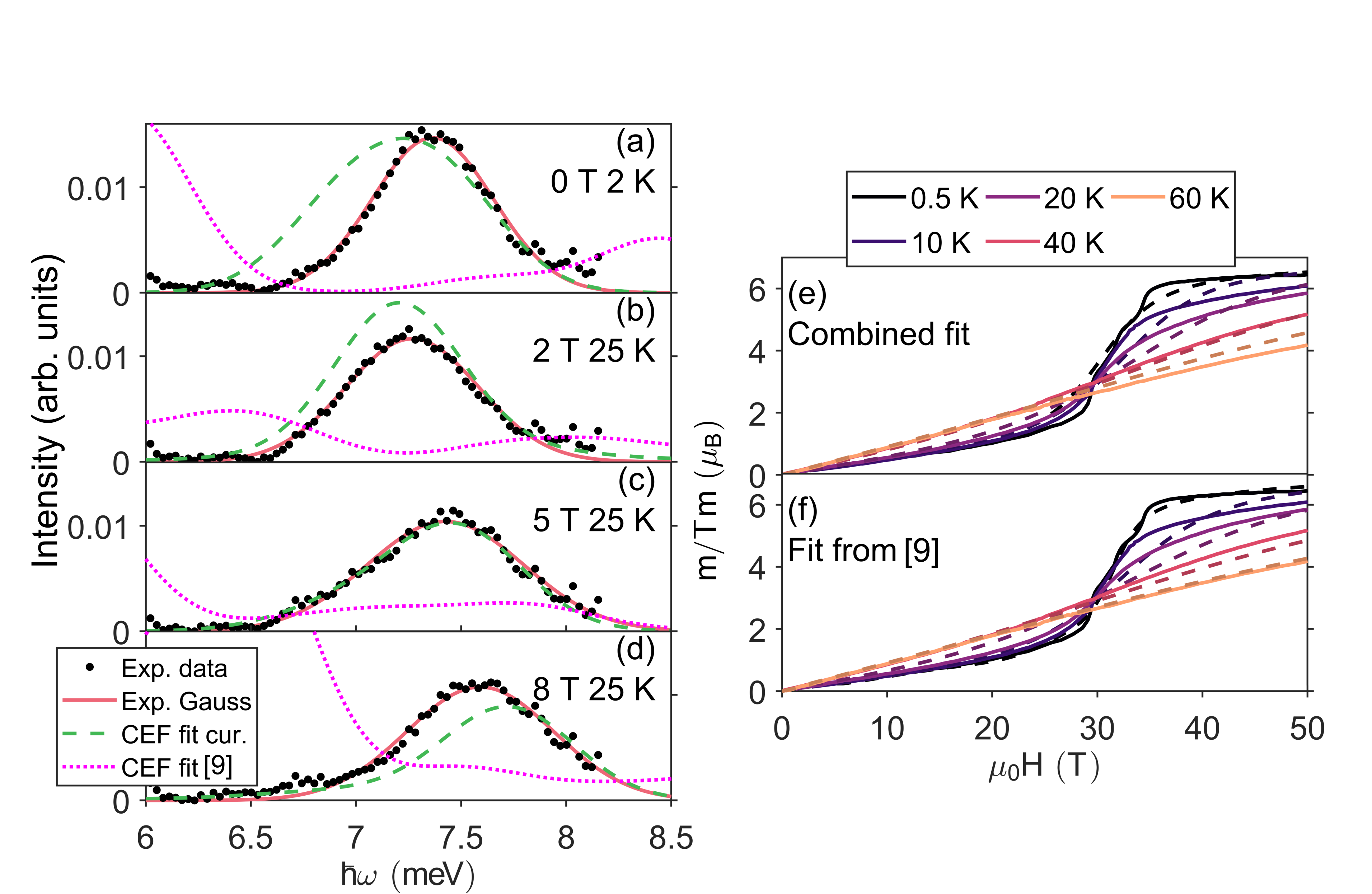}
        \caption{(a)-(d) INS data from CAMEA (black dots) collected under four different field and temperature conditions. The experimental intensity cuts were obtained by integrating over Q along l from [0 3/2 -1/2] to [0 3/2 1] over the observable hk0 range to obtain quasi-powder intensity profiles. The red line are Gaussian fits to the experimental data, which were used for the CEF determination. The remaining curves are calculations based on our combined CEF refinement and the CEF scheme reported in Ref.~\citenum{goddard2007}. (e)-(f) Bulk magnetization data from Ref.~\citenum{morosan2005} (solid lines), presented alongside predictions of the refined CEF scheme (dashed lines).}
        \label{fig:cef_fit_results}
    \end{figure}

    After integration of the neutron scattering data over the observable ($0 k l$) range, we incorporated them into a combined fit with the $H \| c$ magnetization data which were weighted with a cost function:
    \begin{equation}
         R^2_\text{cost} = R^2_\text{INS} + \alpha R^2_\text{mag}.
    \end{equation}
     $R^2_\text{INS}$ and $R^2_\text{mag}$ are the residuals of the INS and magnetization fits respectively, and $\alpha$ is a relative weighing factor. For the presented combined fit, a weighing factor of 2\% was used as this was found to result in a rough agreement with both data sets. The fits and subsequent CEF calculations were carried out with the PyCrystalField python package~\cite{scheie2021}. CEF parameters calculated from a point charge model as implemented in Ref.~\citenum{scheie2021} were used as starting parameters. The resulting fits are shown in Fig.~\ref{fig:cef_fit_results} alongside predictions that are based on the reported CEF. The INS data provide complementary insight into the CEF model of the system. This is evident, for instance, by the fact that the purely magnetization-derived parameters fail to accurately account for the INS results. In contrast, our combined fit displays a reasonable agreement with both data sets. It should be noted, that while we believe our fit improves upon the CEF model of Ref.~\citenum{goddard2007}, it is still an approximate model due to the few peaks in the TmAgGe neutron spectra and the known difficulty of fitting CEF parameters with limited data~\cite{scheie2022un}.

    \begin{table}[]
        \centering
        \begin{tabular}{l| c c}
        \hline \hline
        \diagbox{meV}{Fits}      &  \textbf{Combined fit} & \textbf{Ref.~\citenum{goddard2007}} \\ \hline
        $\mathbf{B_2^0}$ & $-5.44\cdot10^{-2}$ & $1.12\cdot10^{-1}$ \\
        $\mathbf{B_2^2}$ & $-8.16\cdot10^{-1}$ & - \\
        $\mathbf{B_4^0}$ & $-6.57\cdot10^{-4}$ & $-2.67\cdot10^{-4}$ \\
        $\mathbf{B_4^2}$ & $-1.05\cdot10^{-4}$ & $1.60\cdot10^{-3}$  \\
        $\mathbf{B_4^4}$ & $5.70\cdot10^{-3}$ & - \\
        $\mathbf{B_6^0}$ &  - & - \\
        $\mathbf{B_6^2}$ &  - & - \\
        $\mathbf{B_6^4}$ & $1.69\cdot10^{-5}$ & - \\
        $\mathbf{B_6^6}$ & $1.11\cdot10^{-5}$ & - \\ \hline \hline
        \end{tabular}
        \caption{Crystal field parameters with values $\ge 10^{-6}$ from our combined fit alongside the results from Ref.~\citenum{goddard2007}.}
       \label{tab:Bvals}
    \end{table}

    \begin{table*}[]
     \centering
     \begin{tabular}{l| c c c c c c c c c c c c c c} \hline \hline
     \textbf{Eigenvalue}     & \multicolumn{13}{l}{\textbf{Eigenvectors}} \\
     \textbf{E} (meV) & $|-6\rangle$ &  $|-5\rangle$ &  $|-4\rangle$ &  $|-3\rangle$ &  $|-2\rangle$ &  $|-1\rangle$  &  $|0\rangle$ &  $|1\rangle$ &  $|2\rangle$ &  $|3\rangle$ &  $|4\rangle$ &  $|5\rangle$ &  $|6\rangle$\\
      0.0000 &  0.012 &   & -0.142 &  & -0.491 & & -0.691 &  & -0.491  &  & -0.142 &  & 0.012 \\
      0.0094 &  &  -0.041 &  & -0.307 &  & -0.636 &  & -0.636 &  & -0.307 &  & -0.041 &  &  \\
      6.9733 & -0.278 & & -0.448 & & -0.471 & & & & 0.471 & & 0.448 & & 0.278 \\
      7.4044 & & -0.293 & & -0.561 &  & -0.316 & & 0.316 & & 0.561 & & 0.293 & &  \\
      10.538 & 0.607 & & 0.343 & & 0.029 &  & -0.161 &  & 0.029 &  & 0.343 &  & 0.607\\
      13.502 & -0.621 & & 0.031 & & 0.337 &  &  &  & -0.337 &  & -0.031 &  & 0.621 &  \\
      15.679 & & -0.555 & & -0.379 & & 0.219 & & 0.219 & & -0.379 & & -0.555 &  \\
      23.585 & 0.355 & & -0.517 & & -0.138 & & 0.420 & & -0.138 & & -0.517 & & 0.355 &  \\
      25.061 & & 0.614 &  & -0.140 & & -0.322 &  & 0.322 & & 0.140 &  &  -0.614 &  \\
      38.171 & 0.193 & & -0.546 & & 0.406 & & & & -0.406 & & 0.546 & & -0.193 &   \\
      38.441 & & -0.436 & & 0.512 & & -0.219 & & -0.219 & & 0.512 & & -0.436 &   \\
      55.189 & -0.076 & & 0.308 & & -0.489 & & 0.566 & & -0.489 & & 0.308 & & -0.076 &   \\
      55.208 &  & -0.193 & & 0.408 & & -0.545 & & 0.545 & & -0.408 & & 0.193 &  \\ \hline \hline
      \end{tabular}
        \caption{Eigenvalues and wavefunctions of $|\mathcal{J}\rangle$ for the TmAgGe CEF scheme determined from measurements at CAMEA with $H \| a$ and the magnetization measurements from Ref.~\cite{goddard2007} with $H \| c$.}
     \label{tab:CEFscheme}
     
 \end{table*}

    $B_2^2$ is the dominant term in our combined fit model (cf.~Table~\ref{tab:Bvals}). This is similar to what has been observed in the sibling compound YbAgGe, where the magnetic site also bears a strong $B_2^2$-type crystal field~\cite{matsumara2004}. TmAgGe thus exhibits similar anisotropic properties than YbAgGe, The anisotropy is, however, more extreme in TmAgGe which is reflected in the anisotropic g-tensor:
    \begin{equation}
        g = \begin{pmatrix} 13.97 & 0 & 0 \\ 0 & 0 & 0 \\ 0 & 0 & 0 \end{pmatrix}.
    \end{equation}
    The  g-tensor has been calculated~\cite{scheie2021}  on the basis of the CEF scheme of our fit, Table~\ref{tab:CEFscheme}, where the two lowest eigenmodes are treated as a ground state doublet due to their small splitting. The anisotropic g-tensor highlights the Ising-like nature of TmAgGe, where each moment site has its own local quantization axis. We make note of the overall symmetry properties of the eigenstates in Table~\ref{tab:CEFscheme}. The ground state quasidoublet consists of two non-magnetic singlet states, which combine to give rise to a magnetic moment with an easy axis along the local x-direction~\cite{gubbens1998}. The calculated CEF wave-functions are also consistent with the lack of observed low energy modes. The spin dynamics of an Ising system is expected to correspond to localized moment flips. These longitudinal excitations are invisible to neutron scattering as all the transition dipolar matrix elements are zero.

\section{Monte Carlo simulations}{\label{Sec5}}

    The field-induced magnetic moment arrangements of TmAgGe determined from single crystal neutron diffraction allow us to extend the modeling of the reported magnetic exchange Hamiltonian of Ref.~\citenum{goddard2007}. We performed classical Monte Carlo simulations using a minimal effective magnetic Hamiltonian for a single layer of the distorted kagome lattice.  MC simulations using the Metropolis algorithm combining standard temperature dependent changes of the magnetic moments plus local spin flips (Ising like moves) and overrelaxation (microcanonical) updates were used to increase the acceptance rate in the anisotropic local environment. We used an annealing scheme to lower the temperature at fixed external magnetic fields. Simulations were performed for $3\times L^2$ sites ($L = 12-48$) under periodic boundary conditions. 10$^5$ -10$^6$ MC steps were used for an initial relaxation, and measurements were taken in twice as many MC steps for $20$ independent realisations. The effective Hamiltonian is given by
    \begin{equation}
        \begin{split}
        \mathcal{\hat{H}}&=\mathcal{J}^2\sum_{ij}J_{ij}{\bf \hat{S}}_{i}\cdot{\bf \hat{S}}_{j}+\mathcal{J}^2A\sum_{i}(\hat{n}_{i}\cdot{\bf \hat{S}}_{i})^2\\[1ex]
        &-g_J\,\mathcal{J}\,\mu_0\mu_B{\bm H}\cdot\sum_{i}{\bf \hat{S}}_i.
        \label{eq:Hamil}
        \end{split}
    \end{equation}

    The first term corresponds to the Heisenberg bilinear exchange couplings $J_{ij}$ between sites $i$ and $j$ in the layer, $\bf S$ is a unitary spin vector. We considered couplings up to the third nearest neighbour (Fig.~\ref{fig:phd} (b)), $i.e.$ $J_{3a}$ and $J_{3b}$ terms were added to the Hamiltonian of Ref.~\citenum{goddard2007}. The second term in the Hamiltonian is the axial local single-ion anisotropy. The last term is a conventional Zeeman term of a magnetic field ${\bm H}$. The total angular momentum quantum number of the Tm$^{3+}$ ion is $\mathcal{J}=6$ and the Land\'e factor $g_J=7/6$. 

\begin{figure}[htb]
        \includegraphics[width=86mm]{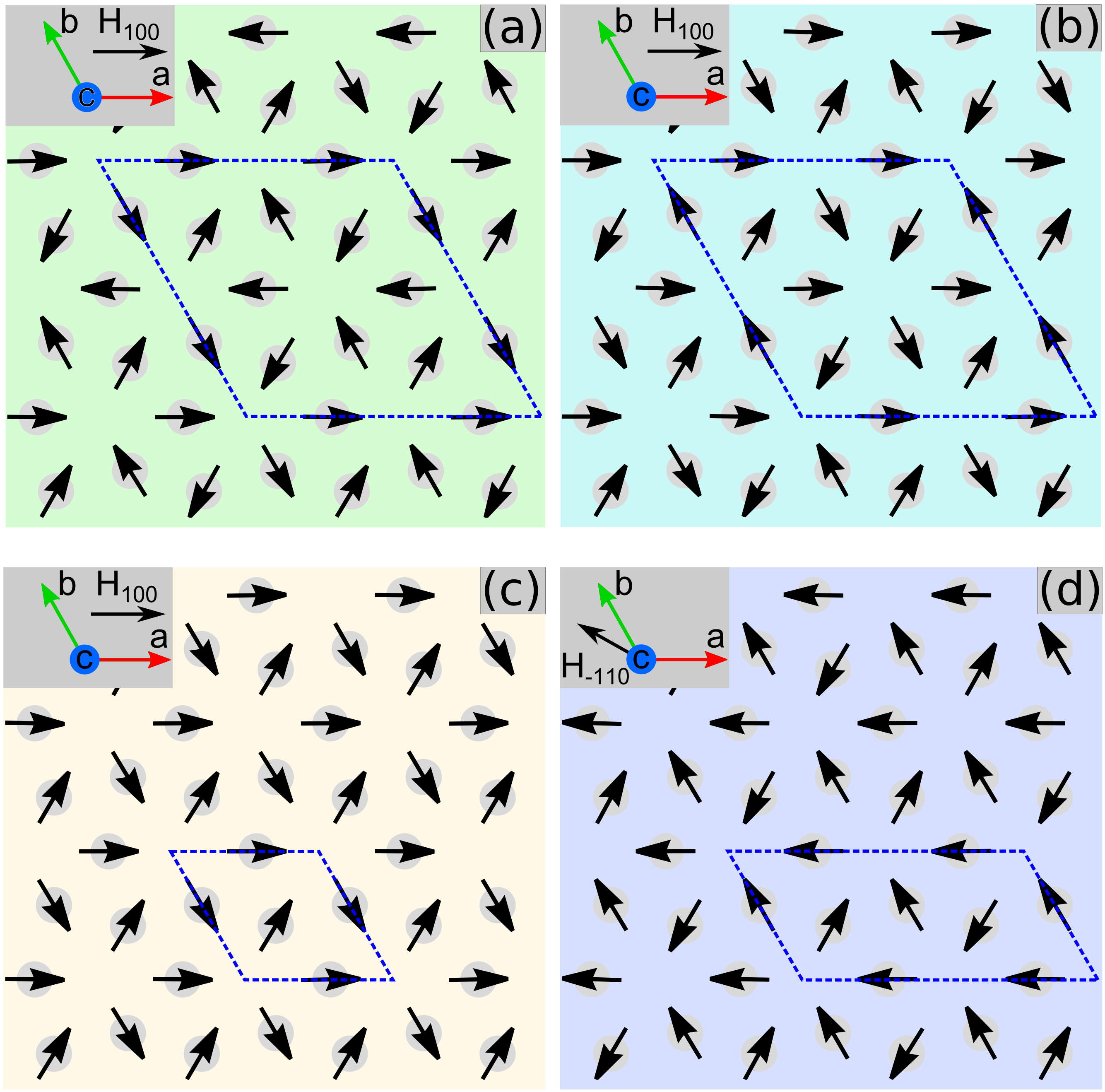}
        \caption{Spin configurations for each state obtained from MC calculations to determine the exchange couplings. For $H//[100]$, (a) $M/M_s=0$, (b) $M/M_s=1/3$ and (c) $M/M_s=2/3$. For $H//[-110]$, (d) $M/M_s=1/\sqrt{3}$.}
    \label{fig:snapsNS-MC_sim}
    \end{figure}

    Using magnetic configurations determined by SND (Fig.~\ref{fig:snapsNS-MC_sim}), we computed the exact energy of the model in Eq.~\ref{eq:Hamil} as a function of applied magnetic fields at zero temperature. The critical fields associated with the transitions between the different magnetic structures were calculated from the relationships between the exchange constants:

    \begin{eqnarray}
    \label{eq:Bc1}
        H^{M0-M2}_c&=&4\,J_{3b},\quad H_{\|[100]}\\
    \label{eq:Bc2}
        H^{M2-M4}_c&=&J_1+\frac{1}{2}J_2+8\,J_{3b},\quad H_{//[100]}\\
    \label{eq:Bc3}
        H^{M0-M3}_c&=&\frac{1}{\sqrt{3}}(J_1+\frac{1}{2}J_2+8\,J_{3b}),\quad H_{//[-110]}
    \end{eqnarray}

    Here, $M_0$, $M_2$, $M_4$ and $M_3$ correspond to the arrangements shown in Fig.~\ref{fig:snapsNS-MC_sim} for $M/M_s=0$ (panel (a)), for $M/M_s=1/3$ (panel (b)), for $M/M_s=2/3$  (panel (c)) and for $M/M_s=1/\sqrt{3}$ (panel (d)), respectively. The first three arrangements match the multi-$k$ SND solutions presented in Fig.~\ref{fig:refinedstruc}  (a-d), while for the last arrangement we used a simplified equal-moment approximation of the SND solution from Fig.~\ref{fig:refinedstruc} (e).

\begin{figure}
\includegraphics[width=86mm]{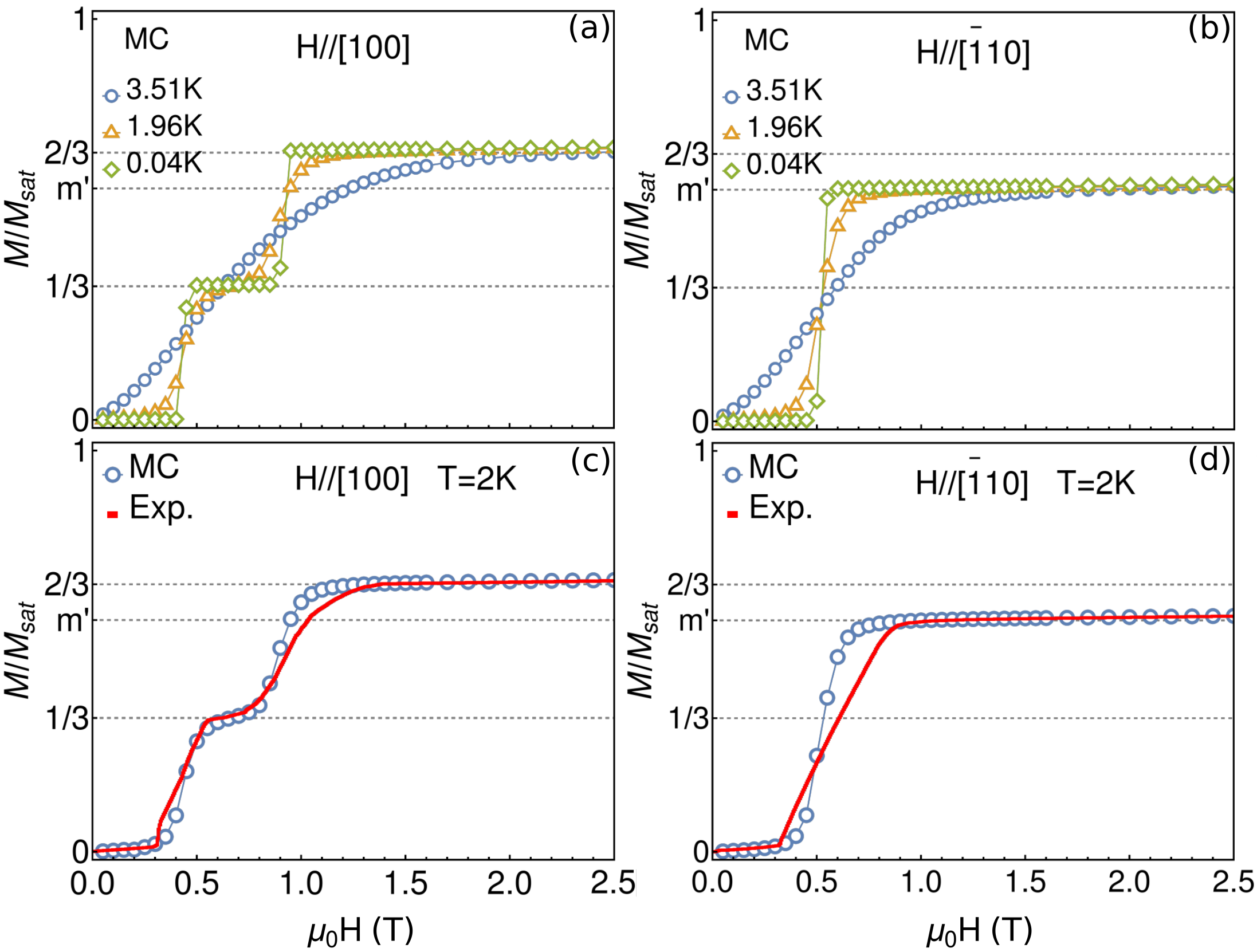}
\caption{ $M/M_s$ predicted by our model for $H\|$[100] (a) and $H\|$[[-110] (b) at different temperatures. Panels (c) and (d) show a comparison with the experimental data at $T=2$ K. $m'$ equals to $1/\sqrt{3}$.}
\label{fig:MvsB}
\end{figure}

We notice that i) the critical fields do not depend on $J_{3a}$ and ii) that Eqs.~(\ref{eq:Bc2}) and (\ref{eq:Bc3}) are not independent. Therefore, the equations were solved for the critical fields as function of one exchange coupling, $i. e.$ $J_2$ defines the energy scale. The sign and size of $J_{3a}$ has consequences on the existence of a $M/M_s=1/6$ energy plateau, whereas  $J_2$/$J_1$ influences the presence of a $M/M_s=1/2$ plateau. As these plateaus were not observed experimentally, energies configurations with $M/M_s=1/6$ and $M/M_s=1/2$ have to be much higher than the $M/M_s=1/3$ configuration. These extra conditions result into the set of inequalities
    \begin{eqnarray}
        \label{eq:Bc1b}
        2\,J_{3a}&<&J_{2},\quad H_{//[100]} \\
        \label{eq:Bc2b}
        -2\,J_{1}&<&J_{2},\quad H_{//[100]}.
    \end{eqnarray}
\begin{center}
\begin{table}
\begin{tabular}{|c|c|c|c|c|c|c|} 
\hline
&$J_1$ & $J_2$ & $J_{3a}$ & $J_{3b}$ & $A$ &$J_c$\\[1.25ex] 
\hline
value (K)& -0.0348 & 0.075 & -0.07 & 0.0143 & -5.55& -0.134 \\[1.25ex] 
distance ({\AA})& 3.653 & 5.121 & 5.543 & 5.543 & &4.17\\[1.25ex] 
\hline
\end{tabular}
\caption{\label{table:Js} Determined $J_{1,2,3}$ exchange couplings of the model in Eq.~\ref{eq:Hamil} given in units of $\mathcal{J}^{-2}$ and corresponding distances between the Tm$^{3+}$ ions. $\mathcal{J}=6$ is the total angular momentum quantum number for the Tm$^{3+}$ ion. $J_c$ is estimated from the RPA analysis (see section \ref{Sec6}) and $A$ is set by comparing the magnetization curves from MC and experiments.}
\end{table}
\end{center}
A comparison of the critical fields calculated with Eqs.~(\ref{eq:Bc1b}) - (\ref{eq:Bc3}) and the experimental values from Ref.~\citenum{goddard2007} leads to the exchange parameters listed in Tab.~\ref{table:Js}. The calculated magnetization curves are presented in Fig.~\ref{fig:MvsB} matching the experimental data very well.

\begin{figure}
\includegraphics[width=86mm]{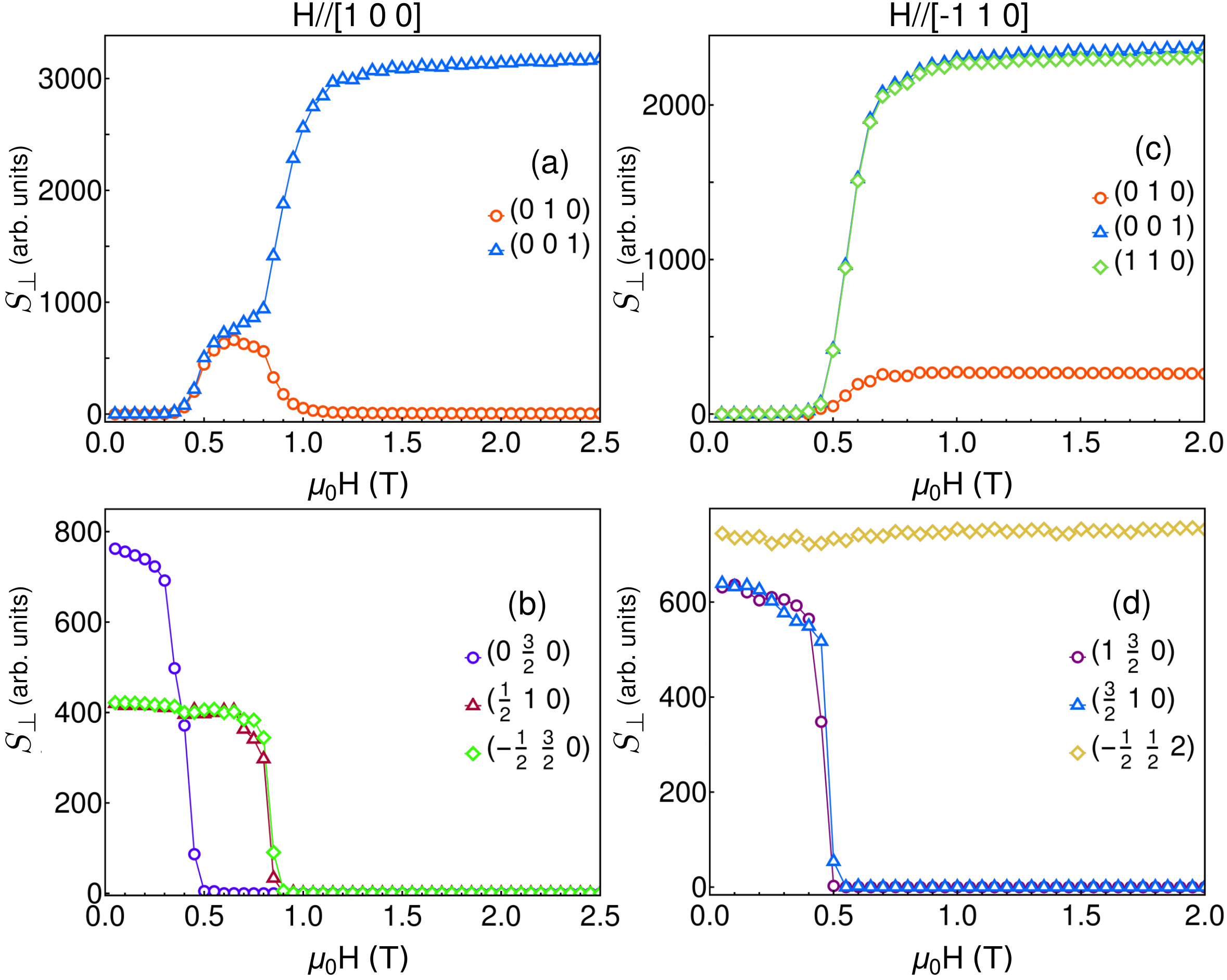}
\caption{Calculated magnetic structure factors for selected $\bf{k}_0$ and $\bf{k}_{1-3}$ reflections as a function of magnetic field applied along ${\bf H}//[100]$ (a, b) and  ${\bf H}//[-110]$ (c, d), respectively.} 
\label{fig:SqvsBcalc} 
\end{figure}

We further computed the static magnetic structure factor $\cal S_\perp$:
\begin{eqnarray}
\cal S_\perp({\bf q})&=&\frac{1}{N}\sum_{a,b}\sum_{i,j=1}^{N}\left(\delta_{ab}-\frac{q^aq^b}{q^2}\right)e^{i{\bf q}\cdot({\bf r}_j-{\bf r}_i)}S^a_{i}\,S^b_{j}
\end{eqnarray}
where $N$ is the number of sites and $a,b=1,2,3$ correspond to the three components of magnetic moment. Fig.~\ref{fig:SqvsBcalc} predicts the field dependence of $\cal S_\perp({\bf q})$ for several $\bf{k_0}$=(0~0~0) and $\bf{k}$=({\half}~0~0) reflections measured in the SND experiment (Fig.~\ref{fig:fielddep}). Also here, a good agreement with the experimental results (Fig. \ref{fig:phd}) is found. 
As $|J_1|$ is smaller than $|J_2|$ we additionally tried to extend the number of parameters by including additional couplings with exchange paths of comparable distances to those already present in our model. These additional exchange couplings beyond $J_{3a}$ and $J_{3b}$ did however not improve the agreement between MC simulations and experimental data and have therefore not been included in our final analysis to avoid overparameterization of our effective model.

We note that the single-ion anisotropy should be large compared to the in-plane exchange couplings, but its precise value cannot be determined by our MC simulations. Similarly we could not determine the out-of-plane coupling $J_{c}$ as we simulated one distorted kagome layer only. 

\section{Random phase approximation}{\label{Sec6}}

Random phase approximation (RPA) calculations were performed to characterize the inelastic neutron scattering results. Most notably, the CEF level around 7 meV remains dispersive within various magnetic states as shown in Fig.~\ref{fig:refinedstruc} (e-h). While magnetic dispersions are generally expected in a magnetically ordered state, CEF levels can also be dispersive in the paramagnetic phase, as for example in elemental Pr~\cite{jensen1991}. CEF excitations are typically localised, but can propagate due to inter-site couplings. Their propagation is coherent and leads to dispersive modes when the temperature is sufficiently low to have weakly populated excited CEF states and a dominant population in the ground state manifold. In this low temperature regime, the dominantly populated ground state acts as an effective order for the coherent propagation of CEF excitations. 
Magnetic order occurring within the ground state manifold, such as the antiferromagnetic order or the field-polarised state of TmAgGe, can modify the effective order out of which the CEF levels are excited and affect their dynamics. One of the most obvious effects is the folding of a dispersion due to a magnetic unit cell larger than the crystallographic unit cell. In the case of TmAgGe, the experimental results indicate that magnetic orders have only a weak effect on the CEF dispersion, which is in agreement with the RPA calculations presented here. This is consistent with the fact that the antiferromagnetic unit cell is not extended along the c-axis and thus, there is no folding along $l$, the direction exhibiting the dominant CEF dispersion.

We considered the following Hamiltonian in our RPA calculations
\begin{equation}
\hat{\cal H}= \sum_{ij} J_{ij} {\bf \hat{J}}_i \cdot {\bf \hat{J}}_j + g_J \mu_0 \mu_B \sum_i {\bf \hat{J}}_i \cdot {\bf H} + \sum_i \hat{\cal H}^\text{CEF}_i ,
\end{equation}
where ${\bf \hat{J}}$ is the total angular momentum operator. 
Here, the CEF wavefunctions from Table~\ref{tab:CEFscheme} were used as basis functions with their energy defined by $\hat{\cal H}_i^\text{CEF}$. In contrast to the Monte Carlo Hamiltonian, there is no anisotropic term, because together with the total angular momentum quantum number it emerges directly from the CEF wavefunctions. For simplicity, only the ground state  and $\sim7~$meV strong first excited quasi-doublets were included.  
The energy of the excited quasi-doublet was fixed at 7.0~meV. Fixing the excited CEF levels at the precise energies in Table~\textcolor{red}{II} instead leads to qualitatively comparable results but with larger shifts of the average CEF energies with temperature and field. The in-plane coupling parameters were fixed to the ones determined from the Monte Carlo calculations (Table~\ref{table:Js}). An additional interaction term $J_c$ between the nearest neighbours along the $c$-axis is required to generate the dispersion along $l$. This value was adjusted to reproduce the dispersion bandwidth of the experimental data at $T=2$~K and $\mu_0H=0$~T. Calculations were performed in the zero-field ordered state ($M_0$), the paramagnetic state and the field-polarised state ($M_4$). The ground state was obtained through a self-consistent mean field approach, using the experimentally determined magnetic unit cell size and is consistent with the experimental and Monte Carlo results. 

\begin{figure}[h!]
\includegraphics[width=86mm]{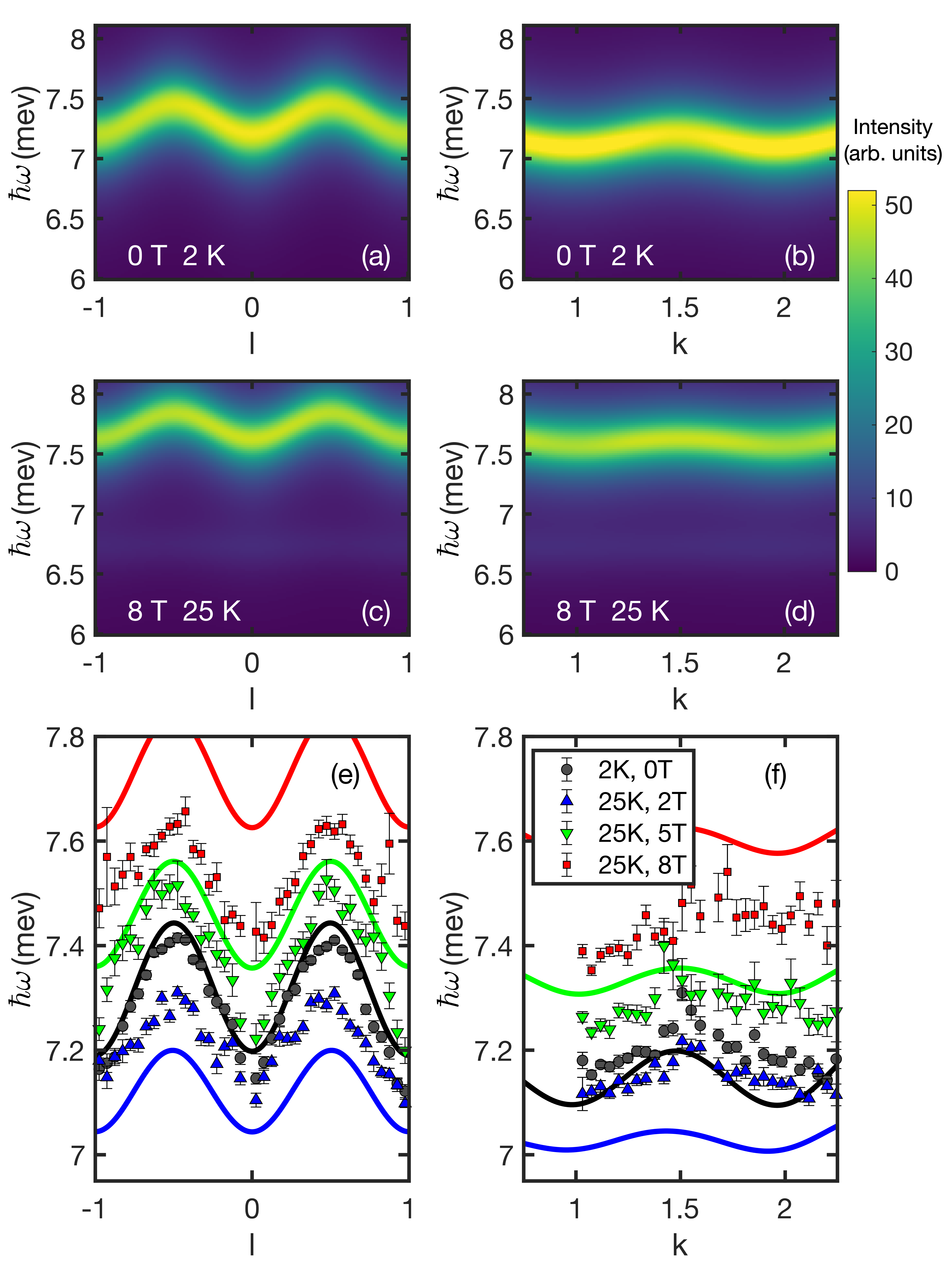}
\caption{(a-d) Dynamic scattering function ${\cal S}_\perp ({\bf Q}, \omega)$ obtained by RPA calculations for cuts along $[0 \frac{3}{2} l]$ and $[0 k 0]$ obtained at field and temperature combinations of 0 T, 2 K and 8 T, 25 K. The results compare well with the corresponding experimental results in Fig.~\ref{fig:cefoverview} (e-h). (e-f) Comparison of the experimental (markers) and calculated (lines) dispersions for various temperature and field values. The experimental dispersions were obtained from Gaussian fits of the experimental spectra such as shown in Fig. \ref{fig:cefoverview} (e-h). Similarly, the theoretical dispersions were obtained via Lorentzian fits of the calculated ${\cal S}_\perp({\bf Q}, \omega)$ in panels (a-d).}
\label{fig:RPAcalc}
\end{figure}

The dynamic scattering function ${\cal S}_\perp ({\bf Q}, \omega)$ obtained by RPA calculations is presented in Fig.~\ref{fig:RPAcalc} (a-d) for the same cuts in reciprocal space, and temperature and field conditions, as the experimental results in Fig.~\ref{fig:cefoverview} (e-h). The calculations reproduce the main characteristics of the experimental data. The experimental dispersion bandwidth along $l$ is 0.25~meV at $T=2~K$ and $\mu_0H=0$~T which is well reproduced by assuming an interaction $J_c=-0.134$~K along the $c$-axis. The dispersion along $k$ has a smaller bandwidth and exhibits a maximum at $k=1.5$, both in the experimental and theoretical results. 

The antiferromagnetic, paramagnetic and field-polarised states were explored by varying the temperature and magnetic field. We find that the overall CEF dispersion is only weakly changed. While at higher temperature the scattering intensity is slightly reduced due to changes in the ground state population, the average energy and bandwidth of the dispersion are also modified. To illustrate this, the extracted experimental and theoretical dispersions are presented together in Fig.~\ref{fig:RPAcalc} (e-f) for various temperatures and fields and exhibit comparable tendencies. We note that our model requires only one adjustable parameter. A better quantitative agreement is expected only, if additional aspects are considered. In particular, the anisotropic dipole-dipole interactions, not included in the model, have an energy scale comparable to the Heisenberg interactions determined by the Monte Carlo calculations. Other possibilities for improvement would be to further refine the CEF wavefunctions or add quadrupolar and higher order multipolar interactions to the model which would create an effective transverse term that leads to propagating flip excitations.

To understand the tendencies observed in TmAgGe, it is instructive to consider a simple RPA model with a ground state singlet $| 0 \rangle$ and an excited singlet $| 1 \rangle$ separated by a gap $\Delta$~\cite{jensen1991}. For a gap much larger than the dispersion bandwidth, the dispersion is given by
\begin{equation}
    E \approx \Delta - p_{01} M_i^2 \mathcal{J}_{ii}(\mathbf{q})
\end{equation}
where $p_{01} = p_0 - p_1$ is the difference in population factor for states $| 0 \rangle$ and $| 1 \rangle$, $M_i=\langle 0|J_i|1\rangle$ is the transition matrix element, and $\mathcal{J}_{ii}(\mathbf{q})$ is the Fourier transform of the couplings. The excited state acquires a dispersion corresponding to $\mathcal{J}_{ii}(\mathbf{q})$ and its bandwidth is scaled by the difference in population factors $p_{01}$ and the square of transition matrix element $M_i$. 
Although TmAgGe hosts a ground state quasi-doublet and an excited quasi-doublet, these simple ideas remain relevant. Thus, the temperature and field dependencies of the bandwidth are associated with changes in the population of these states. A notable example is the bandwidth reduction from the antiferromagnetic ($T=2$~K, $\mu_0H=0$~T) to the paramagnetic state  ($T=25$~K, $\mu_0H=8$~T), an effect well captured by the RPA calculations.

The temperature dependent shift of the average CEF energies stems from the molecular field that is present inside the antiferromagnetic phase and absent in the paramagnetic phase. The shift as a function of field corresponds to the Zeeman effect. Note that the excited doublet splits, but only one level reveals a sufficiently large inter-doublet transition matrix element to be observable in the experimental conditions.

\section{Discussion}{\label{Sec7}}

In TmAgGe the site symmetry and local environment lead to the quasi-doublet ground- and first excited states of the non-Kramers Tm$^{3+}$ ions. Our CEF analysis shows that only the $\hat{J}_x$ dipolar operator attains finite values within each quasi-doublet, because $\hat{J}_y$ and $\hat{J}_z$ are exactly zero due to the symmetry. As a consequence, the Tm$^{3+}$ ground state acts as an exact Ising moment. The experimental evidence for this statement stems from the INS spectra and the combined refinement of the magnetization and INS data.\\
The strong CEF anisotropy and $\pm$ 60$^\circ$ easy axes rotation of adjacent Tm$^{3+}$ moments impose a rigid constraint on the emergent magnetic arrangements, according to the refinements of our SND data. In contrast to the Ho-analogue~\cite{zhao2020}, where the CEF anisotropy imposes the ice rule on the $J_{1}$-triangles, in TmAgGe the ice rule is fulfilled for the $J_{2}$-triangles (Fig.~\ref{fig:refinedstruc}).
In TmAgGe the competition among the in-plane couplings leads to a combination of several propagation k-vectors, $i.e.$ the $\bf{k}$=({\half}~0~0) arms in zero field and the same $\bf{k}$ as well as $\bf{k_0}$=(0~0~0) in applied magnetic fields. The realised magnetic structures are multi-k by the virtue of the single $\bf{k}_{1-3}$ solutions being improbable anisotropic different-moment structures, as our ND results show. We point out that TmAgGe is a rather rare example of such multi-k situation.\\
The zero-field M$_0$ phase hosts a structure with 'one-in-two-out' and 'two-in-one-out' $J_{2}$-triangles. 
Metamagnetic transitions occur via reverse flips of the magnetic moments with orientation adverse to the field direction. For $H \| a$ the flips occur first to the unfavorable Tm1-moments in the M$_2$-phase. Then the unfavorable Tm2- and Tm3-moments flip, setting up the single-k structure with a single 'one-in-two-out' $J_{2}$-triangle in the M$_4$- phase. Field along $H \|[-110]$ forces the unfavorable Tm1- and Tm2- moments to flip, resulting in two 'one-in-two-out' and 'two-in-one-out' $J_{2}$-triangles in the M$_3$ phase. The M$_3$- and M$_4$-arrangements remain stable up to the highest applied fields of 10 T, as they are the best compromises between the local anisotropy axes and the particular magnetic field directions.\\
The moment orientations of the refined structures are in agreement with the magnetization-derived configurations of Ref.~\citenum{morosan2005}, which were based on the triple coplanar Ising model. The magnetisation- and Monte Carlo-derived H$_c$-$\theta$ phase diagrams are very similar. The only discrepancy is the M$_1$ region, which MC results do not identify as a separate phase. This could arise from the different temperatures - 0.04K for the calculation and 2K for the measurement, or due to additional terms, like dipolar interactions, which are not included in our model. Our SND data show a coexistence of the M$_0$ and M$_1$ phases in this region.  \\
We note that in the M$_3$ state the SND refinement suggests two distinct moment values. Here the magnetic moments canted by 30$^\circ$ from the applied field $H//[-110]$ are significantly reduced compared to the moments orthogonal to the field, thus resembling the partially ordered Kagome-ice phase observed in HoAgGe~\cite{zhao2020}.
We speculate that the missing part of the ordered moment is dissipated within domain walls predicted for transverse field Ising models~\cite{sen1992}.\\
Our MC simulations of the distorted kagome layer Hamiltonian allowed determination of all significant in-plane exchange couplings: $J_{1}$= -0.0348 K, $J_{2}$= 0.075 K, $J_{3a}$=-0.07 K, $J_{3b}$=0.01143 K. The ferro- $J_{1}$ and antiferromagnetic $J_{2}$ couplings were already proposed in Ref.~\citenum{goddard2007}. The earlier requirement to suppress one $J_{1}$ coupling is lifted in our extended model through the introduction of the two relevant $J_{3}$ couplings, which feature the same distance between Tm$^{3+}$ ions but follow different exchange paths. In fact, this difference have been shown to be important for the formation of rational plateaux in the kagome lattice~\cite{Albarracin2013}. The different sign of $J_{3a}$ and $J_{3b}$ is essential to stabilize the $\bf{k}$=({\half}~0~0) multi-k structure and for the emergence of the plateaus in applied fields.\\
The measured INS spectra did not reveal any signatures for low-energy excitations stemming from exchange couplings found in the MC simulations that would characterize the ordered magnetic state. 
This is in agreement with the Ising nature of the ordered state anticipating that the spin dynamics corresponds to localised moment flips. These longitudinal excitations are invisible to neutron scattering as all the transition dipolar matrix elements are zero. In contrast, our INS data revealed a transition to the excited quasi-doublet, for which the inter-doublet transition matrix elements for $\hat{J}_y$ and $\hat{J}_z$ are nonzero. These transverse terms, together with the intersite couplings enable the excited CEF states to propagate on the lattice, creating the observed dispersion.
The determined CEF scheme and the in-plane couplings are successfully 
used by the RPA analysis.
With only one adjustable $J_c$ parameter ($J_c$= -0.134 K) we find a reasonable agreement between the calculated and observed spectra. According to the RPA calculations, the $J_c$ interaction is the dominant coupling. This suggests that TmAgGe should be understood as a ferromagnetic Ising chain system that orders in a three-dimensional antiferromagnetic fashion due to weaker in-plane couplings. Yet, these in-plane couplings are responsible for the peculiar field dependence.\\
Interestingly, some of the magnetically ordered phases, $i. e.$ the field-induced M$_3$ and M$_4$ states, are topologically nontrivial, as the total chirality is finite. It remains to be explored whether the topological arrangement of the local magnetic moments affects the electronic band structure of TmAgGe. 
It is conceivable that the observed multi-k magnetic orders in TmAgGe ascribe to the Fermi surface nesting between the M-points and M-$\Gamma$ points.
This would be in line with topological predictions~\cite{barros2014}, suggesting that electronic band structure of the Kagome lattice has a Fermi surface nesting among the M-points. This nesting combined with a divergent density of states at certain fillings makes the system unstable towards development of a triple-k magnetic order. The understanding of the band electronic structure of TmAgGe is crucial to dwell into this speculation.\\
Finally, we anticipate that the presented study facilitates the understanding of other, more complex representatives of the RAgGe family. We find a satisfactory description of the magnetic properties of TmAgGe representing Ising moments that interact via bilinear exchange interactions. Yet we speculate that the localised electrons interact with the itinerant subsystem. While TmAgGe represents a rather localised member of the RAgGe family, YbAgGe we think is located at the crossover where the itinerant electron subsystem gains dominance~\cite{budko2004, niklowitz2006, schmiedeschoff2011, tokiwa2013}. The strong electronic correlations at this crossover is often thought to trigger the emergence of unconventional quantum phases such as quantum spin liquids, hidden orders or unconventional superconducting states. 
\begin{acknowledgments}
This work was performed at SINQ, Paul Scherrer Institute, Villigen, Switzerland with financial support of the Swiss National Science Foundation (Grant Nos. 200020-182536 and 200021$\_$200653). N.G. acknowledges the support of the Canada First Research Excellence Fund (CFREF). H. D. R. and F. A. G. A. are partially supported by CONICET (PIP 2021-112200200101480CO), SECyT UNLP PI+D X893 and PICT-2020-SERIEA03205.  J. L. is supported by the Danish National Committee for Research Infrastructure through DanScatt Grant No 7129-00006B and through a Swiss Data Science Center LSI Track Grant  (project number: C21-16L). Work done at Ames National Laboratory (PCC and SLB) was supported by the U.S. Department of Energy, Office of Basic Energy Science, Division of Materials Sciences and Engineering. Ames National Laboratory is operated for the U.S. Department of Energy by Iowa State University under Contract No. DE-AC02-07CH11358. We acknowledge the long-standing interest of Prof. Ch. R\"{u}egg in this project, useful discussions with S. Petit and A. L\"{a}uchli. We thank M. P. Avicena for help during the CAMEA experiment and I. Plokhikh for help during preparation of the crystals for the ZEBRA experiment.
\end{acknowledgments}

%

\clearpage
\appendix

\section{Symmetry-allowed magnetic space groups}

\begin{figure}[H]
    \centering
    \includegraphics[width = 8.6cm]{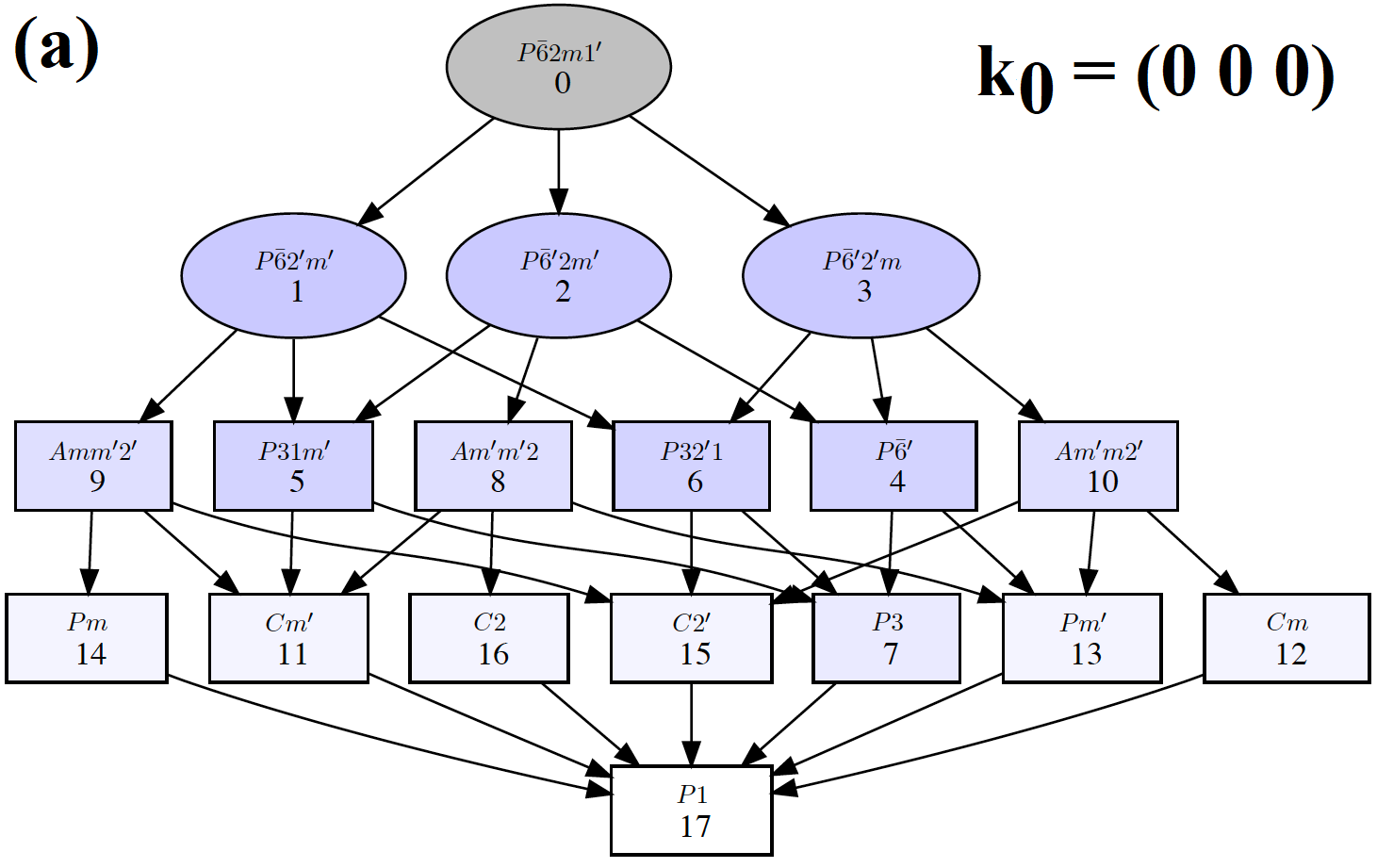}
    \includegraphics[width = 8.6cm]{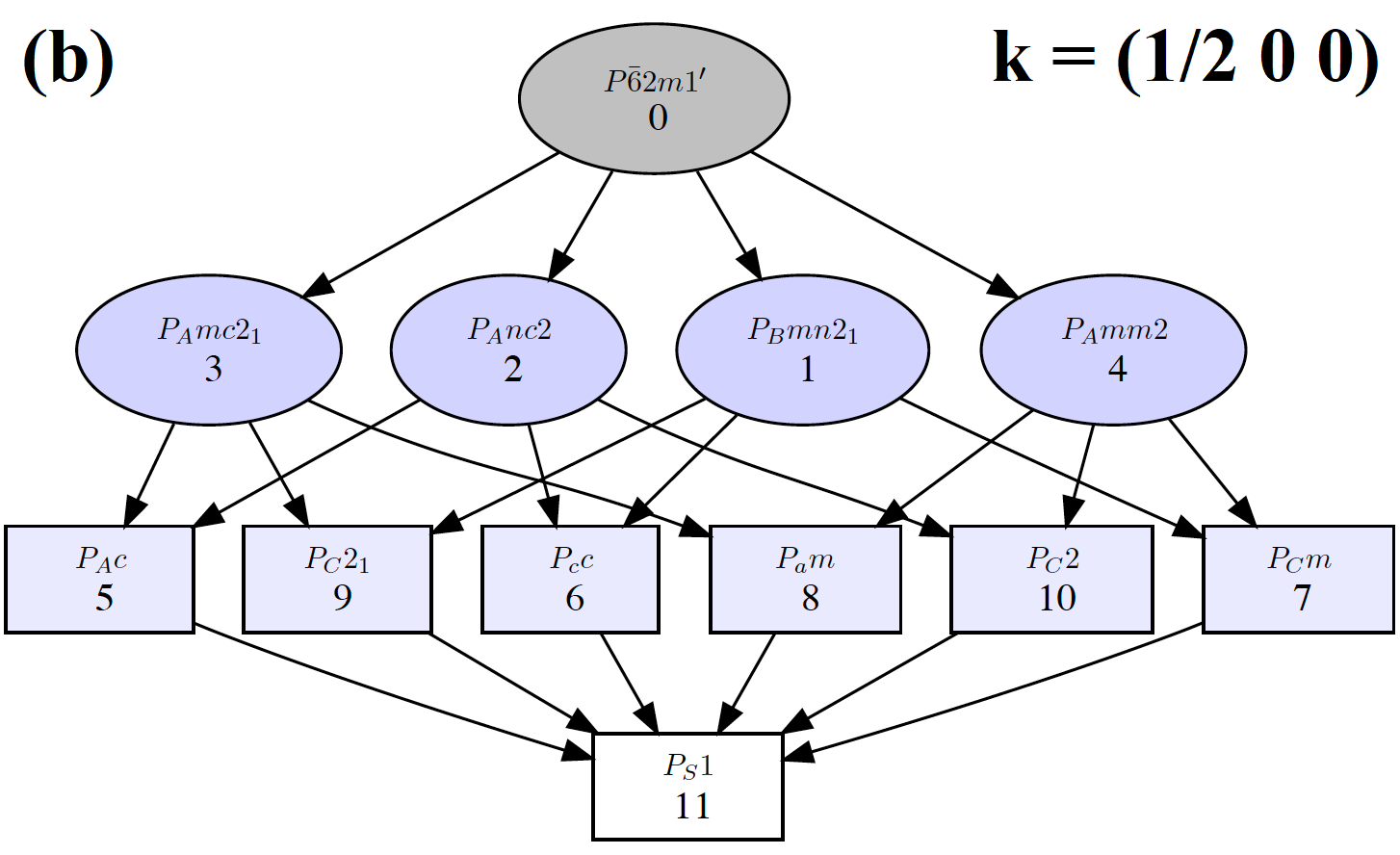}
    \caption{Possible single-$k$ magnetic structure solutions for the (a) $\bf{k}_0$ = (0~0~0) and (b) $\bf{k}$=({\half}~0~0) propagation vector with the $P\bar{6}2m$ paramagnetic parent phase.}
    \label{fig:MSG_singlek}
\end{figure}

    Fig.~\ref{fig:MSG_singlek} lists the possible magnetic space groups that are compatible with the $P\bar{6}2m$ paramagnetic parent phase and a single-$k$ propagation vector that is either $\bf{k}_0$ = (0~0~0) or $\bf{k}$=({\half}~0~0). 
    These lists were derived from the \emph{k-SUBGROUPSMAG} program~\cite{perezmato2015} from the Bilbao Crystallographic server\cite{aroyo2006, aroyo2006v2, aroyo2011} using as input the propagation vector, the paramagnetic space group $P\bar{6}2m$  (\# 189), and the $3g$ Wyckoff position of the magnetic Tm$^{3+}$ ions. Only solutions that allowed finite magnetic moments at all Tm sites were generated. The figures themselves were generated with \emph{SUBGROUPGRAPH}~\cite{ivantchev2000}, also from the Bilbao Crystallographic server, with ellipses indicating $k$-maximal symmetries, while squares are possible subgroups.

\begin{figure*}
    \centering
    \includegraphics[width = 0.9 \linewidth]{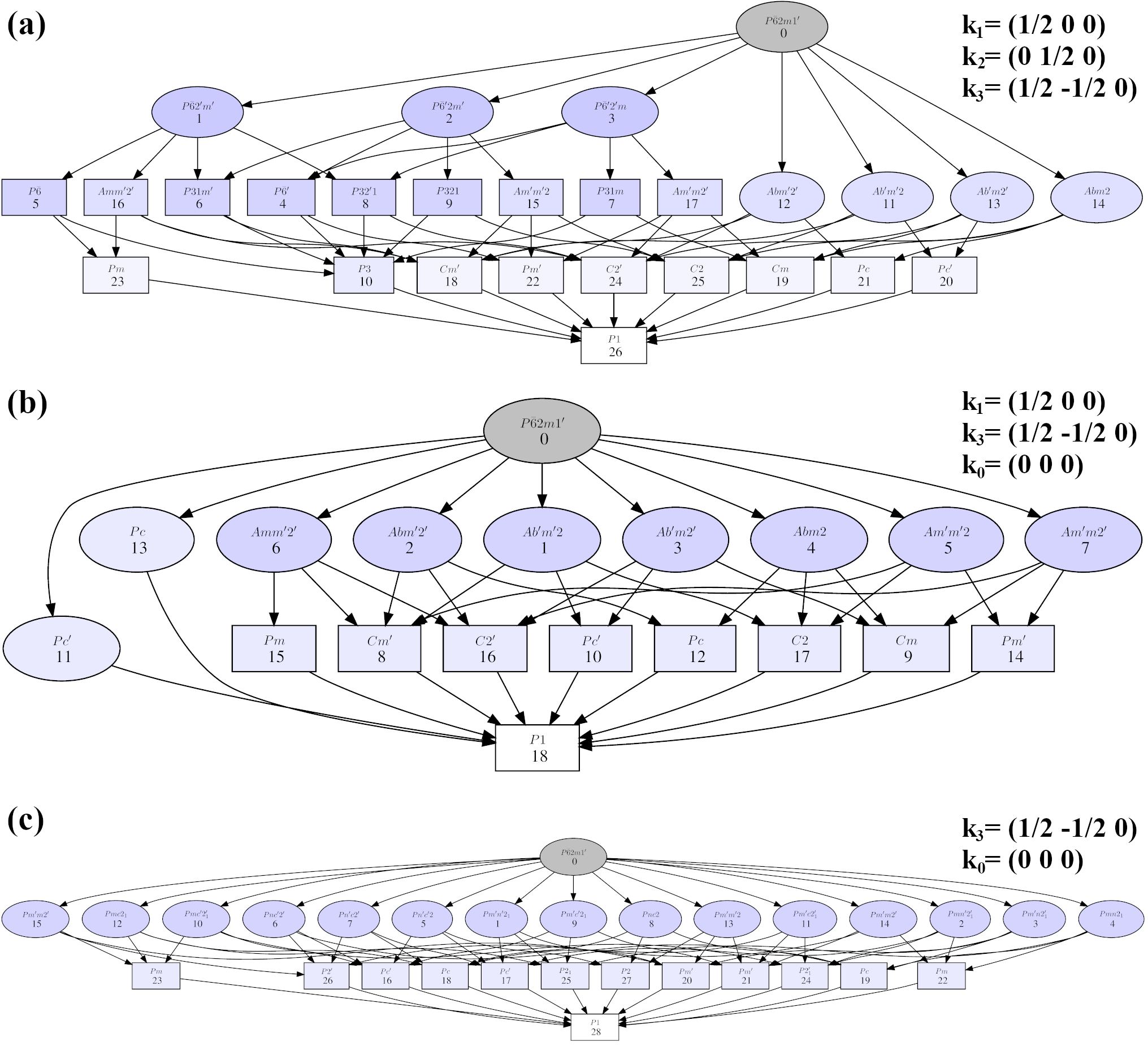}
    \caption{Possible magnetic groups for ordering of the Tm moments in different multi-k cases, corresponding to the magnetic propagation vectors observed at (a) zero field, (b) $H$ along $a$, $\mu_0 H$ = 0.65 T, and (c) $H$ along [-1 1 0], $\mu_0 H$ = 1 T.}
    \label{fig:MSG_multik}
\end{figure*}

    Fig.~\ref{fig:MSG_multik} shows a similar analysis performed for the different possible multi-$k$ structures, based on the combination of propagation vectors observed during single-crystal diffraction experiments under different field configurations.

\section{Magnetic structure refinement results}

A summary of the magnetic structure solutions with the best goodness-of-fit parameters for each collected data set is shown in Table.~\ref{tab:refinement_summary}. For each field configuration, both single-$k$ and multi-$k$ solutions were attempted and were shown to have similar goodness-of-fit parameters. This highlights how diffraction alone cannot be used to decipher between the multi-$k$ and single-$k$ solutions. 

\begin{table*}
    \centering
    \begin{tabular}{l || a{1cm} a{1.5cm} a{1.5cm} | a{2cm} || d{1cm} d{1.5cm} d{1cm} | d{2cm} } \hline \hline
           & \multicolumn{4}{c}{\cellcolor[RGB]{213,252,210} M$_0$ \textbf{ 0 T}}  & \multicolumn{4}{c}{\cellcolor[RGB]{203,247,247}M$_2$~$ \bf{H\| a}$, \textbf{0.65 T}} \\
           &\multicolumn{3}{c}{\cellcolor[RGB]{213,252,210}\textbf{single-k}}&\cellcolor[RGB]{213,252,210}\textbf{multi-k} & \multicolumn{3}{c}{\cellcolor[RGB]{203,247,247}\textbf{single-k}}&\cellcolor[RGB]{203,247,247}\textbf{multi-k} \\
           & $\bf{k}_1$ & $\bf{k}_2$ $\rightarrow$ $\bf{k}_1$  & $\bf{k}_3$ $\rightarrow$ $\bf{k}_1$  & $\bf{k}_1$+$\bf{k}_2$+$\bf{k}_3$ &  $\bf{k}_1$ & $\bf{k}_3$ $\rightarrow$ $\bf{k}_1$ & $\bf{k}_0$  & $\bf{k}_1$+$\bf{k}_3$+$\bf{k}_0$ \\ \hline
           \textbf{N}$_\textbf{eff}$ &  32 & 16 & 44 & 92 & 17 & 42 & 41 & 100\\
           \boldsymbol{$\chi$}$^2$ & 123 & 76.5 & 19.0 & 87.3 & 8.96 & 21.0 & 22.0 & 17.4\\
           \textbf{RF2} & 21.0 & 17.1 & 6.53 & 16.4 & 13.1 & 37.7 & 35.0 & 32.5\\
           \textbf{RF2w} & 23.8 & 17.1 & 11.2 & 22.4 & 14.0 & 35.1 & 37.4 & 30.1\\
           \textbf{RF} & 12.5 & 12.0 & 4.70 & 10.3 &  6.60 & 21.5 & 18.4 & 17.2\\ \hline
           \textbf{m$_1$} ($\mu_B$) & 0.3(2) & 0.9(3) & 0.0(1) & 6.3(1) & 4.9(1) & 4.7(1) & 4.6(2) & 6.7(1)\\
           \textbf{m$_2$} ($\mu_B$) & 6.1(1) & 6.0(1) & 6.8(1) & 6.3(1) & 0.1(3) & 0.4(3) & 4.6(2) & 6.7(1)\\
           \textbf{m$_3$} ($\mu_B$) & 0.3(2) & 0.9(3) & 0.0(1) & 6.3(1) & 4.9(1) & 4.7(1) & 0.2(5) & 6.7(1)\\
           \textbf{m$_H$} ($\mu_B$/Tm) & 0 & 0 & 0 & 0 & 0 & 0 & 1.7(2) & 2.23(3)\\
           \textbf{k}$_\textbf{c}$ & 0 & 0 & 0 & 0 & 0 & 0 & 0 & 0\\ \hline \hline
    \end{tabular} \\ \vspace{0.5cm}

    \begin{tabular}{l || e{3cm} || f{1.5cm} f{1.2cm} | f{3cm} } \hline \hline
      & M$_4$~$\bf{H\| a}$, \textbf{2 T} & \multicolumn{3}{c}{\cellcolor{lightpurple} M$_3$~$\bf{H\| [-1 1 0]}$, \textbf{1 T}} \\
      & \textbf{single-k} & \multicolumn{2}{c}{\cellcolor{lightpurple}\textbf{single-k}} & \textbf{multi-k} \\ 
      & $\bf{k}_0$ & $\bf{k}_3$ $\rightarrow$ $\bf{k}_1$  &  $\bf{k}_0$  & $\bf{k}_3$+$\bf{k}_0$ \\ \hline
      \textbf{N}$_\textbf{eff}$ & 57 & 74 & 35 & 109\\
      \boldsymbol{$\chi$}$^2$ & 30.8 & 6.22 & 12.5 & 7.11\\
      \textbf{RF2} & 21.4  &  7.96 & 30.0 & 9.41\\
      \textbf{RF2w} & 16.0 & 13.5 & 28.7 & 15.0\\
      \textbf{RF} & 13.1 & 5.24 & 15.8 & 6.11\\ \hline
      \textbf{m$_1$} ($\mu_B$) & 6.1(1)  & 0.1(1) & 5.9(2) & 5.3(2)\\
      \textbf{m$_2$} ($\mu_B$) & 6.1(1)  & 7.9(1) & 5.9(2) & 5.3(2)\\
      \textbf{m$_3$} ($\mu_B$) & 6.1(1)  & 0.1(1) & 1.1(3) & 8.2(1)\\
      \textbf{m$_H$} ($\mu_B$/Tm) & 4.09(4) & 0 & 3.0(1) & 3.0(1)\\
      \textbf{k}$_\textbf{c}$ & -1/3  & 0 & 0 & -1/3\\ \hline \hline
    \end{tabular}
    \caption{Summary of \emph{FullProf} refinement results for different field condition data sets. For each of the four different configurations, both single- and multi-$k$ refinements were attempted. The $\bf{k}_n$ $\rightarrow$ $\bf{k}_1$ notation indicates that the data set contains reflections from the $\bf{k}_n$ arm, while the refinement was carried out with the reflections rotated to $\bf{k}_1$, so that the same magnetic space group formulations could be used for all arms of the $\bf{k}_{1-3}$ star. }
    \label{tab:refinement_summary}
\end{table*}

\end{document}